%% file: main.tex
\documentclass[twocolumn,iop,apj,colorlinks,citecolor=blue,urlcolor=blue,linkcolor=blue]{openjournal}
\usepackage{newtx}
\usepackage{placeins}
\usepackage[T1]{fontenc}
\usepackage{natbib}
\usepackage{graphicx}
\usepackage{graphics}
\usepackage{color}
\usepackage{hyperref}
\usepackage{float}
\usepackage{multirow}
\usepackage{booktabs}
\usepackage{orcidlink}
\usepackage[export]{adjustbox}
\usepackage{amsmath}
\usepackage{amstext}
\usepackage{savesym}
\savesymbol{tablenum}
\restoresymbol{SIX}{tablenum}
\usepackage{enumitem}
\DeclareRobustCommand{\rchi}{{\mathpalette\irchi\relax}}
\newcommand{\irchi}[2]{\raisebox{\depth}{$#1\chi$}}

\begin{document}

\title{millimeter-wave Detections of Symbiotic Stars in SPT and ACT data}
\shorttitle{mm-wave Symbiotic Stars}
\shortauthors{Tandoi et al.}
\thanks{Corresponding author: \email{ctandoi2@illinois.edu}}


\begin{abstract}
We present the results of a joint targeted search of candidate symbiotic stars at millimeter wavelengths using the South Pole Telescope (SPT) and the Atacama Cosmology Telescope (ACT). Candidates are selected from the New Online Database of Symbiotic Variables, restricting to objects that are within either the SPT-3G or ACT~DR6 footprint, covering most of the southern hemisphere and up to a declination of $+20^\circ$. Forced photometry on the 828 candidate symbiotic star locations in SPT and ACT data results in 31 unique objects detected with more than a $3\sigma$ significance using two frequency bands: 18 confirmed and 13 suspected symbiotic stars. We provide the SPT and ACT 95/98, 150, and 220~GHz light curves, along with optical and infrared light curves from 2016--2026, as well as spectral energy distributions, physical parameters from the literature, and brief summaries regarding the nature of each individual object. Using Herschel SPIRE data from 2013, we place upper limits on millimeter flux for CN Cha near the beginning of the optical rise in its 2012/2013 nova, which suggests a strong variability and lag at millimeter wavelengths and results in a rare observance of a potential Galactic millimeter slow transient. In addition, we provide coadded thumbnails and light curves for the remaining 797 candidate symbiotic stars that did not pass our detection thresholds. Millimeter-wave emission from symbiotic stars is primarily a combination of free-free emission of the ionization region and optically thick blackbody emission of the cooler dust components of the system. When combined with contemporaneous multi-wavelength observations, millimeter-wave observations can be used to test binary models of symbiotic stars and provide insight on the geometry and physical properties of these systems. 
\end{abstract}

\input{authors}


\section{Introduction} \label{sec:introduction}
Symbiotic stars (SySts) are interacting binary systems of evolved intermediate mass stars: a cool red giant (RG) which donates mass and a hot compact companion which accretes mass \citep{berman32, boyarchuk67, boyarchuk68}, these accretors typically are white dwarfs (WD) \citep{tutukov76} but have also been observed to be neutron stars (NS) in rare instances. These systems serve as valuable astrophysical laboratories for study of the evolution of intermediate mass stars, mass loss and interacting winds, accretion disks and jets, ionization of dense nebulae, complex and irregular variability---including the possibility of being type Ia supernovae (SNe Ia) progenitors---and the physical processes driving these phenomena \citep{mikolajewska12}.

The number of confirmed SySts has been continuously increasing over the years. While different methods of classification are used, recent databases are generally in agreement. \citet{akras19} finds 257 in the Milky Way (MW) and 66 extragalactic, while \citet{merc26} uses a list of 400 SySts: 329 in the MW and 71 extragalactic. Candidate SySts of varying levels of confidence number about twice as many, with \citet{merc19} listing $\sim$850 as ``Likely'', ``Possible'', or ``Suspected.'' These numbers are still much lower than theoretical estimates: \citet{laversveiler25} present multiple methods for estimating the size of the underlying population of SySts, with values ranging from 1.2$\times10^4$--4$\times10^5$ of MW SySts, while previous studies have suggested lower ranges of $\sim3\times10^3$--$3\times10^4$ \citep{yungelson95} and 1.2$\times10^3$--1.5$\times10^4$ \citep{lu06}. The low number of extragalactic SySts can be explained by the already difficult nature of identifying SySts on top of much larger distances; while symbiotic outbursts could be bright enough to see, definitive proof of these is mainly due to lack of observations \citep{ilkiewicz19}. Studies of SySts in the Magellanic Clouds have yielded different population statistics than the MW, possibly due to differing metallicities \citep{mikolajewska04}, which could bias searches. As new SySts continue to be discovered, it is clear that the identification and confirmation of SySts is still a field in development.  

It is likely that nearly all SySts in the Galaxy have been \textit{detected} already in the $K$ band, as red giant branch (RGB) and asymptotic giant branch (AGB) stars tend to have absolute magnitudes of $M_K=-7$ mag or brighter, which corresponds to $K=14.5$ mag even with a $K$-band extinction of $A_K=5$ mag and a distance of 20 kpc. Between the Galactic Plane surveys undertaken by the United Kingdom Infrared Telescope \citep{2008MNRAS.391..136L} and the European Southern Observatory's VISTA Variables in the Via Lactea project \citep{2010NewA...15..433M,2012A&A...537A.107S,2024A&A...689A.148S}, the whole Galactic Plane has been surveyed to several magnitudes fainter than this value; only in the most reddened or crowded regions are SySts likely to have gone undetected.  For objects outside the Galactic Plane, the 2MASS limits typically reach $K_S=14.3$ mag \citep{2006AJ....131.1163S}, and sources should be at quite low extinction, so only the most distant parts of the halo could have SySts not detected in existing surveys.  Furthermore, WISE also covers nearly the whole sky to roughly $W1=16.6$ mag \citep{cutri12}, except in crowded regions, so out of the Galactic Plane, nearly all SySt should be detected with WISE.

The major challenges lie in identifying which objects are symbiotics among the much larger population of AGB and RGB stars. Wide-field surveys in wavelengths where SySts emit but relatively few other giants emit, like millimeter (mm), thus hold promise for helping to grow samples of such objects, and an investigation of the detection rates of already-known SySts represents the first step in evaluating this technique for identifying which objects merit spectroscopic follow-up.

Recently, SPT and ACT have published studies on mm emission of nearby objects including flaring stars \citep{naess21, guns21, tandoi24, wan26, biermann25}, asteroids \citep{chichura22, orlowski24}, and satellites \citep{foster25}. While these have been observations of exclusively transient objects---either changes in luminosity or changes in position---this paper uses similar pipelines and adds to the mm studies of MW objects using cosmic microwave background (CMB) experiments.

The organization of this paper is as follows: 
In Section~\ref{sec:systs} we provide more information about SySts and their classifications, as well as mm emission mechanisms in SySts and different models that utilize them. In Section~\ref{sec:data} we describe the New Online Database of Symbiotic Variables (NODSV) catalog, the South Pole Telescope (SPT), Atacama Cosmology Telescope (ACT), methodology for performing this search, and use of external data. In Section~\ref{sec:discussion} we discuss the results of our analysis.  In Section~\ref{sec:conclusion} we summarize and conclude. In Section \ref{sec:appendix} we provide information for a small population of non-detected D- and D'-type SySts. In Section~\ref{sec:individuals} we give a brief history in the literature of each detected SySt along with any important information for that system, as well as SEDs and multi-wavelength light curves.


\section{Symbiotic Stars} \label{sec:systs}
 In this section we provide basic summaries for information relevant to the context of this paper.  
 
\setcounter{footnote}{0}

\subsection{Classifications}
Characteristics in the infrared (IR) spectral energy distribution (SED) of the giants are the primary classifier for SySts, and generally fall into three IR types:
\begin{itemize}[align=right]
    \item \textbf{S-type (stellar)}: The donors in S-type SySts are typically first-ascent RGB stars, though recent studies have made a strong case for AGB stars also being the donors in some of these systems \citep{merc25b, boffin25}. Their SEDs peak at shorter wavelengths ($\sim$ 1$\mu$m) than do the other classes, and are dominated by the photospheres of the giants, which are usually of spectral type M. These are the most populous type of SySt detected, at $\sim$80\% in existing catalogs \citep{belczynski00, akras19}.
    \item \textbf{D-type (dusty)}: Usually the donors in these systems are AGB stars and often Mira variables. Their SEDs peak at higher wavelengths ($\sim$2-4 $\mu$m) as the dust shells they are surrounded by can obscure the M-giant photosphere. These make up the majority of the remaining SySt types at $\sim$15\% \citep{akras19}.
    \item \textbf{D'-type}: These are a distinct class of dusty SySts, showing a much cooler SED peak that tends to be flat around $\sim$10-30 $\mu$m, and they contain F/G/K type giants. These make up $\sim$5\% of detected SySts \citep{akras19}.
\end{itemize}

Most S-type SySts have M-type cool components, with some having warmer K- or G-type giant donors within the system instead. Together with D'-type SySts these GK giant donor systems are known as yellow symbiotics \citep{glass73, allen82}. S-type yellow symbiotics are not obscured by dust shells and the absorption continuum of the giant can be observed. They can show an overabundance of s-process elements, but are not luminous enough to have undergone the third dredge-up phase of an AGB star's evolution. They thus must have accreted those elements from the current accretor in the system when it was in its own AGB phase, and are therefore important in studying the history of mass transfer in the binary \citep{pereira17}. Although S-type yellow symbiotics are quite rare, with 12 known in the literature \citep{baella16}, we make note of them due to multiple detections we present in Section~\ref{sec:individuals}.

As mentioned above, recent interferometric observations of a sample of 13 confirmed S-type SySts have yielded updated radii estimates which, when combined with reddening-corrected Gaia magnitudes and placed on evolutionary tracks of stars with typical mass and metallicty ranges for S-type SySts, indicate that most of this population is likely already in the AGB phase--though not yet evolved into a Mira variable--or at least at the tip of the RGB. These results suggest that deriving parameters for these stars based on typical, non-symbiotic giants of similar spectral types may not be valid \citep{merc25b, boffin25}.

\subsection{Outbursts and Variability}
The outbursts of SySts are related to the accretion by the WD beyond its quiescent phase, where it is releasing energy at a relatively constant rate, perhaps with some modest stochastic variability.  In Z~And outbursts the accretion rate surpasses the threshold for stable burning on the surface of the WD, which leads to an outburst on the timescale of weeks to years, and can be prolonged by multiple re-brightenings \citep{skopal20}. There is ambiguity behind the mechanism responsible with different possibilities suggested, including an increase of mass transfer rate from the RG (either through intrinsic variability or enhanced capture of the wind near periastron), an enhanced wind leading to the formation of an optically thick disk around the WD or a disruption of the accretion disk from the enhanced wind, or some change in the colliding winds between the WD and RG \citep{munari19}. 

For SyN the mechanism is more clear. Accretion of matter on a typical WD with mass of $\sim$0.4-0.8 M$_\odot$ \citep{mikolajewska07} can eventually lead to hydrogen burning on the surface of the WD in non-degenerate conditions. SyN are non-explosive and happen in thermal equilibrium, with the thermonuclear burning envelope expanding to massive sizes and continuing to burn in stable conditions. SyN exhibit slow outbursts that can go on for decades, with the outbursts of lower mass WDs taking longer to decline \citep{mikolajewska10}. 

As these novae depend on a mass accretion threshold and do not irreversibly disrupt the system, all novae are potentially recurrent: those with recurrence periods on short timescales ($<100$ yrs) and with multiple observed outbursts are known as symbiotic recurrent novae (SyRN) \citep{darnley21, merc25c}. SyRN are related to SyN in that they also result from thermonuclear ignition on the surface of the WD due to accretion. However, these occur on the surfaces of massive WD ($\sim$1.2-1.4 M$_\odot$) where the accreted material is degenerate. Thermonuclear ignition can be triggered on the surface of the WD in a process similar to the classical novae seen in binaries where a WD accretes from a red dwarf (RD) companion. The shell does not expand as the temperature increases due to the degenerate state, until finally the temperature reaches a critical level, the electron degeneracy is lifted, and a violent outburst occurs.  Where they differ is in what happens after the outburst: classical novae (which represent the same basic nuclear fusion process, but in systems with dwarf or subgiant donors) see the WD shell get expelled from the system which will then expand freely, with some internal shocks taking place due to variations in the shell speeds over time (e.g. \citealt{2020NatAs...4..776A}). In SySts the circumstellar medium (CSM) is formed by dense RG winds and the nova dumps energy into the nebula as it is decelerated, leading to shocks and emitting energy across the entire electromagnetic spectrum \citep{munari25}.

While the novae themselves are not irreversibly disruptive, there has been speculation that SySts can be SNe Ia progenitors. In the single degenerate case SNe Ia happen when the WD accretes enough mass to reach the Chandrasekhar limit ($\sim$1.44 M$_\odot$). The double degenerate case where a double WD merger results in a supernova is also a possibility, as SySts already contain one WD with a companion giant that is potentially headed towards that state of evolution \citep{mikolajewska13}.  One of the core challenges for single degenerate models is that over much of the available parameter space, the mass loss from classical novae may approach or even exceed the mass accreted between novae \citep{1991ApJ...367L..19N}, so understanding the novae in these systems is crucial for assessing the viability of single degenerate channels for Type Ia SNe.

SySts can also show features in their light curves that are related to either intrinsic variability of the RG or due to the orbit of the binary. Some examples are: pulsations in the RG (Mira in D-types and semi-regular in S-types), eclipsing and ellipsoidal effects depending on the inclination of the system as well as reflection effects where the WD heats the RG/nebula, and a temporary increase in accretion rate during periastron passage. Detailed descriptions of these different forms of variability can be found in \citet{gromadzki09, gromadzki13}.

\subsection{Identification of SySts and distinguishing them from imposters}\label{sec:syst_v_pne}
While SySts represent a specific evolutionary path of intermediate mass binary systems, their signature at first glance of ``hot source in cold gas'' is not unique and can lead to confusion with objects such as H\,{\sc ii} regions and, much more commonly, planetary nebulae (PNe) \citep{ilkiewicz17}. 

Formation of PNe occurs at the end of an intermediate mass star's life as the core has finished burning helium and becomes unstable, shedding its outer layers into a rich nebula and leaving only a WD in its wake. To complicate matters in distinguishing SySts and PNe, PNe have their own zoo of classifications \citep{balick02} which can show evidence of central binary stars by having bipolar nebulae \citep{corradi95}, a combination of emission lines and absorption spectra \citep{lutz77}. D'-type SySts in particular are difficult to distinguish (and are often misclassified), as they show great overlap in their spectra with peculiar PNe \citep{schmid93}.

The main difference between SySts and PNe comes down to the components of the system and their interactions: PNe are the remnants of a dead star whose WD core will cool and fade as the nebula expands and dissipates into lower densities, while SySts are actively interacting binaries that often show a range of variability and outbursts as the WD continues to accrete mass inside the higher density nebula. To untangle the two similar objects, careful study of their spectra is necessary. A wide range of spectral lines from high-excitation elements (e.g. [Fe\,{\sc vii}] $\lambda\lambda$ 5721, 6087, He\,{\sc ii} $\lambda4686$, etc.) or indicating high electron density (e.g. [Fe\,{\sc ii}] $\lambda4923$, [S\,{\sc ii}] $\lambda6731$, [N\,{\sc ii}] $\lambda5755$, etc.) give evidence of the WD inside of a nebula. At optical and infrared wavelengths, the photosphere of the giant dominates, showing absorption lines of molecular bands such as TiO, CN, CO, and VO, as well as atomic lines such as Ca\,{\sc i}, Ca\,{\sc ii}, and Fe\,{\sc i} \citep{kenyon87, kenyon88, belczynski00}. 

The Raman-scattered O\,{\sc vi} doublet is the hallmark diagnostic for SySts as they uniquely contain the components necessary for its creation. O\,{\sc vi} $\lambda\lambda$ 1032, 1038 lies very close to Ly$\beta$ ($\lambda$1025.7) and as such can interact with neutral hydrogen \citep{schmid89}. Raman-scattering of the O\,{\sc vi} photon excites hydrogen from its ground state to a virtual level which then emits a photon of $\lambda$6830 or $\lambda$7088. This phenomena requires FUV emission in a dense region to ionize oxygen five times (close proximity to WD) and a steady influx of neutral hydrogen (RG wind) \citep{shore14}. Not all SySts show O\,{\sc vi} Raman-scattered emission lines, with numbers found to be roughly 50\% \citep{allen84a}; however, all detected cases of emission in this line have come from SySts -- with investigations into non-SySt detections resulting in either weak, ambiguous detections of O\,{\sc vi} or reclassification of the object into a SySt \citep{akras19}. The RAMan Search for Extragalactic Symbiotic Stars (RAMSES II) presented a concept for using O\,{\sc vi} filters on already existing telescopes to search for extragalactic SySts \citep{angeloni19}. 

While these are the historical methods for SySt identification, other methods are being explored as new multi-wavelength and time-domain surveys have come online in recent years \citep{merc25}. \citet{merc22} notes that historical methods may be biased towards active, shell-burning WD (as opposed to accreting-only), as most SySts were discovered using low-resolution objective prism photographic surveys which were only capable of detecting SySts with strong emission lines. Additionally, many SySts have significant reddening at optical wavelengths due to their presence in or near the Galactic Plane as well as intrinsic dust reddening in D- and D'-type systems, which bias optical detections further. As dust reddening falls off with increasing wavelength, FIR, mm/sub-mm, and radio observations may be able to probe undiscovered systems better than traditional methods.

\subsection{mm emission}

mm-wave emission observed in SySts is typically a combination of optically thick blackbody emission and free-free radiation (FF) \citep{ivison92}. To describe these mechanisms we use the spectral index $\alpha$, where the flux density at a frequency $S_\nu$ scales with the frequency $\nu$ in the form of $S_\nu \propto \nu^\alpha$. Optically thick blackbody emission ($\alpha=2)$ is modeled with the temperature and size of the emitting region and is dominated by dust shells or nebulae at mm wavelengths, if present, with minimal contributions by the photosphere of the cool giant.

\citet{seaquist84} present a simple binary model for SySts (referred to as STB) in which the radio emission originates as FF from a region of the RG wind that is photoionized by the hot component. The FF turnover frequency ($\nu_t$), where the optical depth $\tau_{\textrm{FF}}$=1, relates to the geometry of the ionization region, the electron temperature, and the binary separation. Optically thin FF emission generally has a spectral index of $-0.1$, while optically thick FF can have a spectral index between 0.6 and 1.3 \citep{taylor84}. 

For D- and D'-type systems, orbital periods are long enough ($>15$ yr) that $\nu_t$ generally falls in the 8-10~GHz range \citep{ivison95b, mikolajewska02}. For S-types, where orbital periods are shorter ($\sim$0.5--15 yr) and $\nu_t$ is in the high GHz/THz regime, direct measurements of the optically thin emission have typically not been possible \citep{mikolajewska02}. Instead, optical and UV spectroscopy is used to approximate the optical depth by measuring H$\beta$ emission.

As mentioned in Section \ref{sec:syst_v_pne}, current diagnostics for distinguishing SySts from PNe require spectroscopy, which, for high extinction sources in the Galactic Plane, can be too time-consuming to be practical or is outright impossible. Radio and mm emission have negligible effects from extinction and when combined can provide an alternative diagnostic by probing the densities of nebulae in candidate SySts, as the densities in D- and D'-type SySt nebulae are generally quite a bit larger than those in PNe.\footnote{Very young PNe can have similar densities \citep{ruiz24}~until they have sufficiently aged and expanded in size.} \citet{2021MNRAS.503.2887B} show that given a radio SED and $\nu_t$, one can estimate the angular size of a FF emitting nebula, and in turn the emission measure. As catalogs of PNe are developed from SPT and ACT, as well as from Southern radio surveys, it will become possible to develop samples of both objects and test whether radio plus mm spectra combined are useful diagnostics.

The STB model gives a general description for the geometry of the ionizing region, however it does not always accurately explain the observed radio/mm emission. The dynamics inside of these complex systems can lead to localizations of various shock fronts due to the high-density, high-velocity winds from both the RG and the WD, and colliding-wind models (CW) are used to explain these, often complementary to STB \citep{girard87, kenny05, kenny07}.

Evidence of non-thermal emission in the form of synchrotron radiation has been shown during the novae of e.g. V1535~Sco \citep{linford17}, V445~Pup \citep{nyamai21}, V3890~Sgr \citep{nyamai23}, and RS~Oph \citep{nayana24}. These are explained as strong shocks as the novae are accelerated into the thick winds of the RG and are observed as radio spectral indices steeper than that of optically thin FF ($-0.1$) that vary in the days following the outbursts. There is also evidence that synchrotron radiation is produced by the jets observed after outbursts \citep{dickey21}.

In non-outbursting SySts, \citet{angeloni07} use simulations of a CW model for H~1-36 to explain the SED from radio through UV. An excess of flux seen at $\sim10^{11}$ Hz  cannot be explained by the various blackbodies (the giant photosphere/dust shells as well as the reprocessed radiation of dust) nor the FF emission from the reverse/expanding shocks. They find optically thin synchrotron radiation, with $\alpha = -0.75$, to be responsible for this excess and can be explained by acceleration of particles due to the Fermi mechanism within the shock front. A similar, though less pronounced, case is found for the quiescent phase between two active phases in 1996-1997 for CH Cyg with the same model \citep{contini09}. Analysis of eight other SySts in \citet{angeloni10} does not find any other synchrotron contributions.

To accurately assess STB and other mm emission models, contemporaneous multi-wavelength observations are necessary as variability and outbursts can affect the entire electromagnetic spectrum: e.g. before and after a nova in CN Cha \citep{lancaster2020}, Mira pulsation periods in the IR spectrum of RR Tel \citep{jurkic12}, and radio measurements of R Aqr \citep{gregory74, bowers79}.


\begin{figure*}[htbp]
    \includegraphics[width=\textwidth,center]{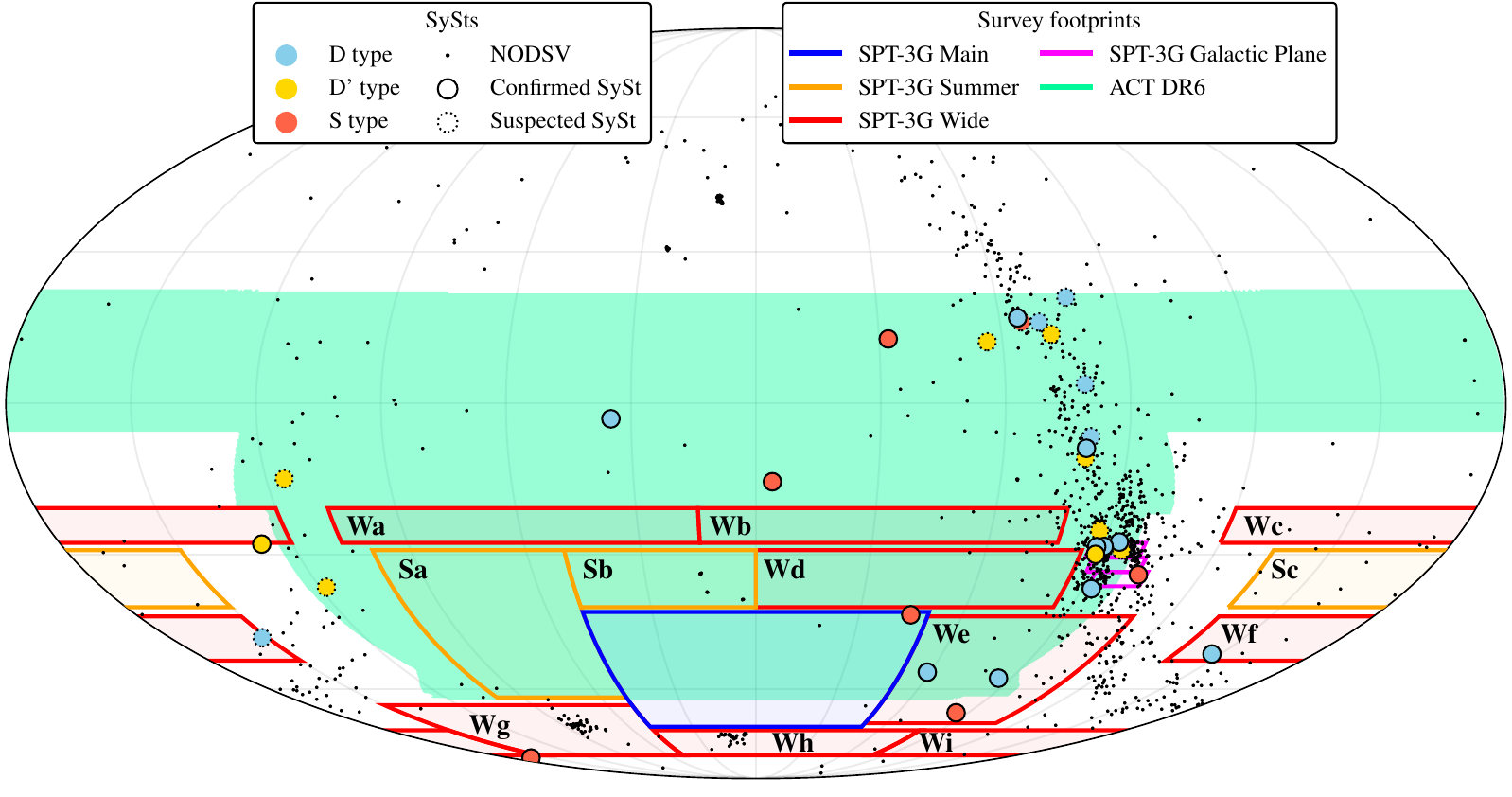}
    \caption{SPT and ACT observing fields, along with all SySts in the NODSV catalog. SPT/ACT detected SySts are shown as colored circles: D-type as light blue, D'-type as yellow, and S-type as red. Candidate status is designated by the circle border: solid lines for confirmed SySts and dotted lines for suspected SySts. The map is centered at Right Ascension=0$^\circ$, decl.=0$^\circ$  with grid lines in 30$^\circ$ increments.}
    \label{fig:skyplot}
\end{figure*}

\section{Data} \label{sec:data}
In this section we describe the data used to detect SySts at mm wavelengths as well as external datasets and any changes made in our tables relative to the literature. Figure~\ref{fig:skyplot} shows the nominal locations of the different SPT fields on the sky, the ACT~DR6 footprint, and all candidate SySts in The New Online Database of Symbiotic Variables (NODSV). 

\subsection{The New Online Database of Symbiotic Variables}
NODSV\footnote{\url{https://sirrah.troja.mff.cuni.cz/~merc/nodsv/index.html}} is an online database of known and suspected SySts with observational data aggregated from the literature \citep{merc19}. As of June, 2025, NODSV contains 1213 total candidate SySts. Information includes alternate identifiers and inclusion in any symbiotic catalogs, astrometry and E(B-V) reddening, multi-wavelength photometry and symbiotic features/diagnostics such as outbursts and emission lines, parameters of the binary system and the individual cool/hot components, and a brief summary of the star with references. Additionally, SySts are labeled with their candidate status: ``Confirmed'', ``Likely'', ``Possible'', etc. with references used to determine the parameters given.

\subsection{South Pole Telescope}\label{sec:spt}

The South Pole Telescope (SPT) has a 10-meter primary mirror and is located at the Amundsen-Scott South Pole Station \citep{carlstrom2011}. SPT-3G has been operational since 2017, and is used to primarily observe the CMB in the southern sky in three frequency bands centered at 95, 150, and 220~GHz with arcminute-scale beams \citep{sobrin2022}. 

The data used in this work come from maps in the extended-10k (Ext-10k) survey: 10,000 deg$^2$ coverage of most of the southern hemisphere between $-20^\circ$ and $-80^\circ$ declination, apart from the Galactic Plane. It is a combination of different observing fields beginning in 2019 \citep{vitrier25}. We also include data from the 2023 and 2024 seasons of the 100 deg$^2$ Galactic Plane survey \citep{wan26}. 

These maps have data binned into 0\farcm25 pixels in a Lambert azimuthal equal-area projection and are filtered using the SPT transient-detection pipeline, i.e. map space filtering consisting of convolutions of an isotropic high-pass filter in the form of a $\sim$3$\arcmin$ annulus to reduce atmospheric noise at large angular scales, and a Gaussian beam template to smooth the maps and maximize point source sensitivity. 

In the transient-detection pipeline, maps in each band are converted from CMB temperature fluctuation units ($T_{\textrm{CMB},\nu}$) to flux density units ($S_\nu$) using estimated conversion factors for each band ($C_{\nu}$). These conversion factors are in units of mJy K$^{-1}$ sr$^{-1}$, and are scaled by the map resolution $\theta$:

\begin{equation}
    S_\nu = T_{\textrm{CMB},\nu} C_{\nu} \theta^2 .
\end{equation}

As the flux densities were calculated using estimations, we adjust them in each band to be consistent with flux densities calculated using a matched filter that includes real SPT-3G beams and bandpasses (see Section 4 in \citealt{archipley26} for a complete description of this method). To do this, we select a population of high SNR point sources in the SPT-3G fields (Archipley, et al. in prep.), apply the transient filter to unfiltered thumbnails of these sources in each band, and compare the matched filtered to transient filtered ratios of flux density for each source and find a linear scaling using a best-fit $\rchi$$^2$. These scalings are then applied to all flux values in their respective frequencies.

In contrast to previous SPT analyses that make use of the transient pipeline, we search in and report flux values from coadded maps of all observations in each field. Light curves of single observation maps (integrated flux values over a $\sim~30$ minute period, with a cadence of $\sim~12$h) have \textit{not} been coadd-subtracted and still include foreground noise. Due to significant non-uniformity of noise across the different fields, noise values for all fields are estimated as the RMS of an annulus around each pixel, following the method used in \citet{wan26}.

For more detailed descriptions of SPT mapmaking, we refer the reader to \citet{dutcher21} and \citet{quan26}, and for transient filtering, we refer the reader to \citet{guns21}, \citet{tandoi24}, and Guns et al., in prep.

\subsection{Atacama Cosmology Telescope}

The Atacama Cosmology Telescope (ACT) was a millimeter wave survey instrument that observed from 2008 to 2022, utilizing three generations of receivers.
These surveys made use of a 6-meter primary mirror to map the CMB with a resolution of approximately 1\farcm5 at 150~GHz.
In this work, we use publicly available data from the third-generation instrument, Advanced ACTpol (advACT, \citealt{henderson16}) which operated 2016--2022.
The advACT instrument housed three optics tubes, each terminating at the focal plane on an array of polarization-sensitive pixels, two dichroic mid-frequency (MF) arrays with band centers of $\sim$98~GHz (f090) and 145~GHz (f150) and one monochromatic high-frequency (HF) array with a band center near 225~GHz (f220).
The advACT survey strategy observed most of the sky accessible from the observatory latitude (-22$^\circ$57\arcmin31\arcsec), a total of roughly 19,000~deg$^2$ (\citealt{naess25}, see Figure \ref{fig:skyplot}). 
More precise advACT passbands and bandcenters can be found in \cite{hervias-caimapo24}.

ACT mapmaking and matched filtering is described in detail in \cite{naess25}.
Of particular interest are the ``depth-1'' maps, with a revisit cadence of roughly once every few days and a pixel resolution of 0\farcm5.
These depth-1 maps are made available through the CMB datasets at the National Energy Research Scientific Computing Center (NERSC)\footnote{\url{https://act.princeton.edu/act-dr6-data-products}} and are used in this work to perform time-resolved photometry.
Typically, the single optics tube depth-1 map noise is 30, 50, and 90~mJy at f090, f150, and f220, respectively, with a time resolution of roughly 5~minutes \citep{biermann25}.

We perform forced photometry on the matched filtered depth-1 maps with the \texttt{pixell}\footnote{\url{https://github.com/simonsobs/pixell}} software utility.
Thumbnail cutouts are also used to construct averages of varying-length time windows to look for longer-term variability.
As stated in \cite{biermann25}, sometimes the depth-1 map noise estimate fails to address local noise enhancements.
To account for this and remove poorly determined fluxes, we take the same approach as \cite{biermann25}, whereby the RMS of the SNR map is used to measure Gaussianity, assuming that the local background is noise dominated.
Observations for which the RMS is greater than 1.5 or less than 0.5 are removed; otherwise the noise is normalized such that the RMS of the SNR is equal to one.

\subsection{External datasets}\label{sec:external_data}

Often, NODSV has multiple possible values for its parameters, with distance being the most frequent ambiguously estimated parameter. For the SPT/ACT detected stars in this study we have investigated each one individually and used the most reasonable measurement of distance. Geometric distance is calculated using Gaia parallax and uncertainty along with a direction-dependent distance prior of stellar populations; photogeometric distances are also calculated using the same information along with color and magnitude of the star and another direction-dependent absolute magnitude prior. \citep{bailerjones21}. As SySts can show additional reddening and dust obscuration effects along with a composite spectrum of a cool red giant + hot blue source, we prefer geometric distances over photogeometric to minimize biases.

The Gaia parallax estimates from DR3 have some potential systematic uncertainties, particularly for systems with orbital periods close to one year, because astrometric wobble can be of the same order as their parallax signature.  For very long period systems, the astrometric wobble leads to an error in estimating the proper motion of the binary, but if the astrometric wobble is nearly linear, that will be the only substantial systematic it contributes. The appendix in \citet{merc25b} explores the effects of this wobble in several SySts. We recommend readers re-visit distances for sources of interest after Gaia DR4 is released, as this will have a longer time baseline and will have epoch astrometry measurements that will allow a more careful treatment of the problem, especially for binaries with known orbital periods.  Where used, these distances do remain the best estimate possible at the current time.

\begin{figure*}[ht!]
    \centering
    \includegraphics[width=\textwidth,center]{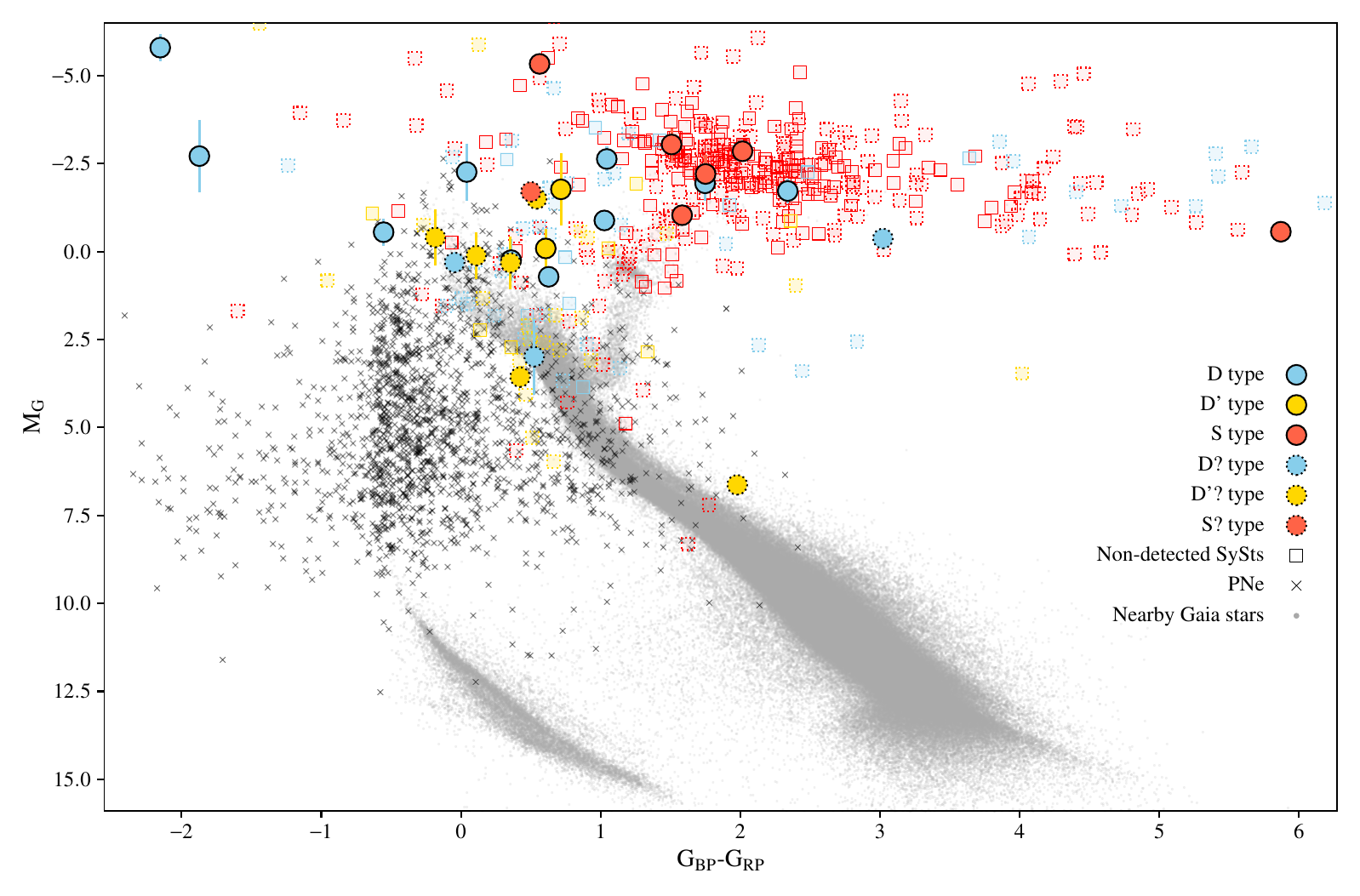}
    \caption{Gaia CMD of the SPT/ACT detected SySts shown alongside nearby ($<$~200~pc) Gaia stars and a population of known PNe \citep{gonzalez21}. Square symbols for non-detected SySts follow the same definitions as circles for detected SySts, with a solid border indicating confirmed SySt and dotted border indicating a suspected SySt. Interstellar reddening and extinction corrections have been applied to the SySts and PNe as described in Section~\ref{sec:external_data}.}
    \label{fig:cmd}
\end{figure*}
Missing data from NODSV we included are IR photometry for three stars: WISE \textit{W1-W4} bands for ShWi~5 and IRAS~18344-0632 \citep{cutri12}, and 2MASS \textit{J, H, K$_{\mathrm{s}}$} bands for WISE~J192140.40+155354.6 \citep{zacharias04}. The S-type SySts R Aqr and CN Cha are both Mira variables and have been previously classified as D-types in the literature \citep{gomez24, lancaster2020}. R Aqr stands out as it has a highly resolved nebula and circumstellar environment and is uniquely well-studied in comparison to other SySts, while CN Cha has recently experienced a nova in 2012/2013 which has changed its SED (see Fig. 2 in \citet{lancaster2020}). Though IR colors for both of these systems support the S-type classification following the diagnostics in \citet{akras19}, we note that this information---along with the previous interferometric studies of S-type SySts in \citet{merc25b, boffin25}---give examples that classification of SySts require accurate measurements and still may blur the line between the different types.

We present a reddening-corrected Gaia color-magnitude diagram (CMD) in Figure~\ref{fig:cmd}. $E(B-V)$ reddening values are supplied in NODSV \citet{schlafly11}, but we follow from \citet{merc26} and calculate new $E(B-V)$ reddening for each SySt using \texttt{mwdust} \citep{bovy16} with the \texttt{Combined15} 3D dust map \citep{drimmel03, marshall06, green15}. We use these values to calculate Gaia $E(B_P-R_P)$ and A$_G$ reddening and extinction values following the relative extinction values provided in \citet{wang19} with R$_{\textrm{V}}$=3.1. We use geometric distances for non-detected candidate SySts to determine their absolute Gaia magnitude $M_G$. D- and D'-type SySts also have additional intrinsic extinction and reddening effects due to the dense dust shells around them which is not accounted for, with the nebula significantly contributing to the Gaia spectrum (see Figure A3 in \citealt{merc26}), and can appear bluer than single giants of the same spectral type. PNe included in Figure~\ref{fig:cmd} come from \citet{gonzalez21}, and also have been corrected for interstellar extinction and reddening effects. 

\begin{figure}[h]
    \centering
    \includegraphics[width=\columnwidth,center]{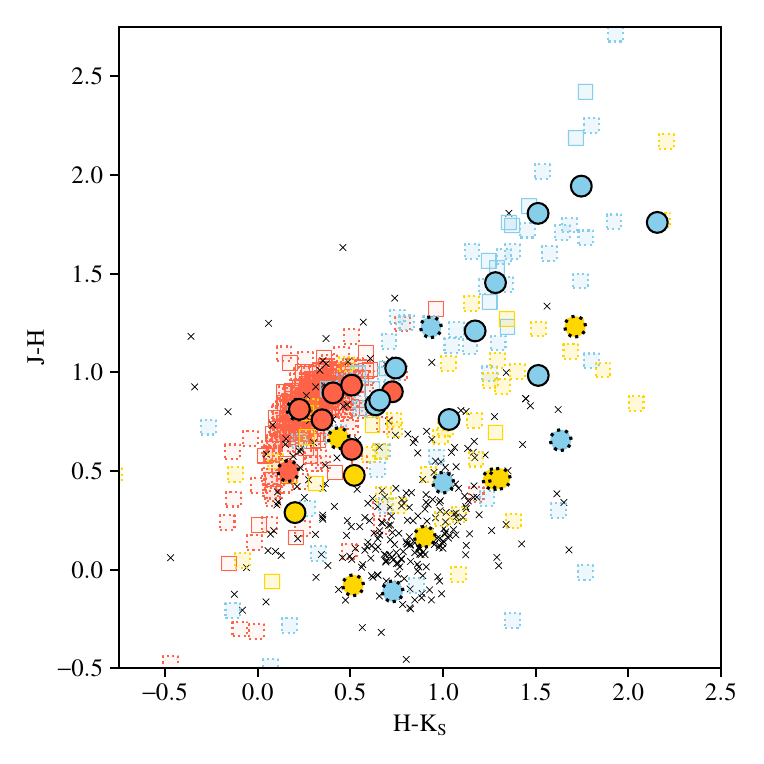}
    \caption{NIR color-color diagram using 2MASS data for all SySts within the SPT and ACT footprints. The markers for SySts and PNe are the same as those used in Figure~\ref{fig:cmd}. Interstellar reddening and extinction corrections have been applied as described in Section \ref{sec:external_data}}
    \label{fig:2mass_color}
\end{figure}

Figure~\ref{fig:2mass_color} shows a NIR color-color plot of all SySts in the SPT/ACT footprints in 2MASS colors, which is a typical diagnostic for separating SySts and PNe. As mentioned previously, Miras can have intrinsic dust reddening leading to much larger NIR color values as shown in Figure 1 in \citet{phillips07}. We also include a population of PNe which come from the Hong Kong/AAO/Strasbourg H$\alpha$ planetary nebula database (HASH) \citep{parker16}. We convert the previously mentioned reddening estimates to the appropriate bands in 2MASS using coefficients provided in Table 2 in \citet{yuan13}: 0.72, 0.46, and 0.306 for $J$, $H$, $K_S$, respectively, and in addition to the WISE data using coefficients of 0.19 and 0.15 for \textit{W1, W2}, respectively. Reddening is negligible at longer wavelengths, so we do not adjust \textit{W3, W4}, or the IRAS data.  

Figures~\ref{fig:cmd} and \ref{fig:2mass_color} highlight that the main confusion between PNe and SySts occurs in D- and D'-type SySts, while S-type SySts tend to exhibit optical and IR characteristics that are different than a typical PNe. This applies to both the SPT/ACT detected and non-detected populations of SySts.

Optical and IR data used for light curves described in Section~\ref{sec:individuals} come from the following: 

Wide-field Infrared Survey Explorer (WISE) light curves \citep{wright10, mainzer11} were acquired from the NEOWISE-R Single Exposure (L1b) Source Table through IRSA\footnote{\url{https://irsa.ipac.caltech.edu/applications/Gator/index.html}} using a 2\farcs5 cone search around the coordinates of each candidate SySt. We use the following data quality cuts after collecting the light curves:

{\small\begin{verbatim}
    qual_frame > 0
    qi_fact == 1.0
    saa_sep > 0
    moon_masked == '00'
\end{verbatim}}
\noindent WISE \textit{W1} and \textit{W2} pixels begin to saturate at 8 and 7 mag, respectively. For sources that saturate the detectors, the point spread function template is enlarged and the fit is performed on non-saturated pixels in the wings. These corrections are provided for magnitudes in the range of $\sim$8-3 for \textit{W1} and $\sim$7-2.5 for \textit{W2}.\footnote{See \url{https://irsa.ipac.caltech.edu/data/WISE/docs/release/NEOWISE/expsup/sec2_1c.html} and \url{https://irsa.ipac.caltech.edu/data/WISE/docs/release/NEOWISE/expsup/sec2_1civa.html} for more details.} We cut observations brighter than the range of corrections offered, as those are not considered reliable. We applied the correction values and their additional uncertainties to the WISE light curves, noting that the NEOWISE Single-exposure Database explains that stars with emission in excess of pure photospheres, such as AGB, can introduce contamination; we have not corrected for this and the resulting photometry may have additional uncertainties. Due to the observing cadence of WISE, data exist as collections of $\sim$10 observations spaced one satellite orbit apart ($\sim$1.6 h) every six months~\citep{petrosky21}. As such we bin these collections together for enhanced visual clarity.

All-Sky Automated Survey for Supernovae (ASAS-SN) optical light curves, using \textit{V} and \textit{g} filters, were obtained using ASAS-SN Sky Patrol\footnote{\url{https://asas-sn.osu.edu/}} \citep{shappee14, kochanek17}. We used the ``Aperture photometry'' method, removed upper limits, and made a cut on data with magnitude SNR~<~5.

Asteroid Terrestrial impact Last Alert System (ATLAS) optical light curves, using ``orange'' and ``cyan'' (\textit{o, c}) filters, were obtained using the ATLAS forced photometry server\footnote{\url{https://fallingstar-data.com/forcedphot/}} \citep{tonry18, smith20}. Using difference images, we followed the data quality cuts they recommend:

{\small\begin{verbatim}
    duJy < 10000 &&
    err == 0 &&
    x > 100 && x < 10460 && y > 100 && y < 10460 &&
    maj < 5 && maj > 1.6 && min < 5 && min > 1.6 &&
    apfit > -1 && apfit < -0.1 &&
    mag5sig > 17 &&
    sky > 17\end{verbatim}}

We note that while the CMD in Figure~\ref{fig:cmd} and the NIR color-color plot in Figure \ref{fig:2mass_color} have had reddening corrections applied, the SEDs and the light curves in Section~\ref{sec:individuals} have not been corrected.


\section{Results and Discussion} \label{sec:discussion}

\begin{table*}
    \footnotesize
    \centering
    \caption{\raggedright\label{tbl:systs_spt}SPT coadded flux of detected candidate SySts with observation date range. This includes both confirmed and suspected SySts.}
    \begin{tabular*}{\textwidth}{l@{\extracolsep{\fill}}rrrr}
    \toprule
    \toprule
    Name  & 95 GHz & 150~GHz & 220~GHz  & Observation Dates  \\
          & [mJy]  & [mJy]   & [mJy]    &   \\
    \midrule
    AS 201 & 4.6 $\pm$ 1.3 & 5.7 $\pm$ 1.1 & 4.9 $\pm$ 4.4 & 1, May 2024 - 19, September 2024 \\
    ASASSN-17dm & 4.3 $\pm$ 2.4 & 7.7 $\pm$ 2.1 & 22.2 $\pm$ 4.7 & 20, January 2024 - 15, February 2024 \\
    CN Cha & 184.4 $\pm$ 19.2 & 220.1 $\pm$ 13.7 & 237.1 $\pm$ 12.5 & 1, January 2024 - 20, January 2024 \\
    V852 Cen & 15.0 $\pm$ 5.7 & 19.7 $\pm$ 3.6 & 25.9 $\pm$ 2.9 & 21, January 2024 - 14, February 2024 \\
    V455 Sco & 5.7 $\pm$ 1.1 & 6.3 $\pm$ 1.8 & 7.3 $\pm$ 6.1 & 13, February 2023 - 20, March 2024 \\
    WSTB 19W032$^*$ & 29.6 $\pm$ 2.0 & 43.7 $\pm$ 3.3 & 82.4 $\pm$ 21.8 & 13, February 2023 - 15, March 2024 \\
    K 5-33$^*$ & 26.0 $\pm$ 3.9 & 28.6 $\pm$ 6.6 & 30.9 $\pm$ 25.8 & 13, February 2023 - 15, March 2024 \\
    H 1-36$^*$ & 48.0 $\pm$ 4.0 & 45.5 $\pm$ 4.2 & 40.4 $\pm$ 5.9 & 21, February 2023 - 16, March 2024 \\
    H 1-45$^*$ & 26.8 $\pm$ 2.8 & 28.6 $\pm$ 3.9 & 46.6 $\pm$ 17.6 & 13, February 2023 - 15, March 2024 \\
    ShWi 5$^*$ & 3.4 $\pm$ 1.1 & 4.4 $\pm$ 1.4 & 9.7 $\pm$ 6.3 & 13, February 2023 - 15, March 2024 \\
    H 2-38$^*$ & 23.0 $\pm$ 2.2 & 25.4 $\pm$ 2.3 & 29.2 $\pm$ 9.9 & 13, February 2023 - 15, March 2024 \\
    Hen 2-375$^*$ & 51.0 $\pm$ 3.8 & 66.8 $\pm$ 3.7 & 82.4 $\pm$ 5.4 & 22, March 2024 - 15, September 2024 \\
    AR Pav & 6.5 $\pm$ 1.3 & 7.6 $\pm$ 1.5 & 7.8 $\pm$ 3.9 & 23, March 2024 - 15, September 2024 \\
    RR Tel$^*$ & 67.4 $\pm$ 4.7 & 76.6 $\pm$ 3.8 & 91.1 $\pm$ 4.0 & 21, March 2024 - 16, September 2024 \\
    DD Mic$^*$ & 3.8 $\pm$ 1.4 & 5.0 $\pm$ 1.1 & 7.7 $\pm$ 1.5 & 25, March 2019 - 26, July 2025 \\
    \bottomrule
    \end{tabular*}
    \begin{minipage}{\textwidth}
        \raggedright
        $^*$: also seen in ACT. DD Mic and ShWi 5 did not meet detection criteria in ACT, but we include their ACT data due to detections in SPT.
    \end{minipage}
\end{table*}

\begin{table*}[ht!]
    \footnotesize
    \centering
    \caption{\raggedright\label{tbl:systs_act}ACT coadded flux of detected candidate SySts with observation date range. This includes both confirmed and suspected SySts.}
    \begin{tabular*}{\textwidth}{l@{\extracolsep{\fill}}rrrr}
    \toprule
    \toprule
    Name  & 95 GHz & 150~GHz & 220~GHz  & Observation Dates  \\
          & [mJy]  & [mJy]   & [mJy]    &   \\
    \midrule
    omi Cet & 37.4 $\pm$ 1.6 & 61.3 $\pm$ 1.9 & 152.5 $\pm$ 7.1 & 25, July 2017 - 1, July 2022 \\
    QX Pup & 52.8 $\pm$ 3.0 & 188.6 $\pm$ 3.6 & 646.6 $\pm$ 14.6 & 1, September 2017 - 26, March 2022 \\
    Sa 2-18 & 33.9 $\pm$ 2.5 & 30.0 $\pm$ 2.7 & 31.0 $\pm$ 9.6 & 7, September 2017 - 30, May 2022 \\
    WSTB 19W032$^*$ & 19.9 $\pm$ 2.8 & 23.9 $\pm$ 2.7 & 24.0 $\pm$ 8.9 & 22, September 2018 - 2, June 2022 \\
    K 5-33$^*$ & 23.2 $\pm$ 3.8 & 23.1 $\pm$ 3.5 & 36.1 $\pm$ 13.4 & 21, May 2017 - 2, June 2022 \\
    H 1-36$^*$ & 55.3 $\pm$ 2.7 & 57.4 $\pm$ 3.0 & 31.9 $\pm$ 10.8 & 21, May 2017 - 28, May 2022 \\
    H 1-45$^*$ & 22.4 $\pm$ 1.8 & 18.4 $\pm$ 1.7 & 13.2 $\pm$ 5.5 & 21, May 2017 - 2, June 2022 \\
    ShWi 5$^*$ & 2.5 $\pm$ 2.2 & 5.0 $\pm$ 2.3 & 14.3 $\pm$ 8.5 & 20, May 2017 - 2, June 2022 \\
    H 2-38$^*$ & 21.3 $\pm$ 2.3 & 28.9 $\pm$ 2.8 & 22.0 $\pm$ 11.0 & 20, May 2017 - 1, June 2022 \\
    MaC 1-10 & 55.5 $\pm$ 2.1 & 58.3 $\pm$ 2.7 & 53.7 $\pm$ 10.9 & 20, May 2017 - 1, June 2022 \\
    Hen 2-375$^*$ & 53.2 $\pm$ 1.6 & 62.9 $\pm$ 1.8 & 67.2 $\pm$ 6.5 & 19, May 2017 - 1, July 2022 \\
    IRAS 18344-0632 & 121.6 $\pm$ 16.3 & 127.8 $\pm$ 6.2 & 60.8 $\pm$ 22.7 & 15, July 2017 - 2, July 2022 \\
    M 1-57 & 32.4 $\pm$ 2.1 & 35.1 $\pm$ 2.5 & 49.2 $\pm$ 9.6 & 15, July 2017 - 2, July 2022 \\
    K 3-9 & 39.0 $\pm$ 2.1 & 47.9 $\pm$ 2.5 & 50.5 $\pm$ 9.7 & 15, July 2017 - 2, July 2022 \\
    PM 1-253 & 20.6 $\pm$ 1.6 & 23.3 $\pm$ 1.9 & 28.0 $\pm$ 7.5 & 16, July 2017 - 2, July 2022 \\
    Hu 2-1 & 85.4 $\pm$ 1.4 & 79.0 $\pm$ 1.6 & 78.9 $\pm$ 6.2 & 14, May 2017 - 2, July 2022 \\
    PM 1-286 & 25.9 $\pm$ 1.4 & 20.1 $\pm$ 1.6 & 7.9 $\pm$ 6.2 & 14, May 2017 - 2, July 2022 \\
    WISE J192140.40+155354.6 & 24.0 $\pm$ 1.5 & 33.1 $\pm$ 1.6 & 48.8 $\pm$ 6.1 & 14, May 2017 - 2, July 2022 \\
    PN Me 1-1 & 30.8 $\pm$ 1.2 & 25.4 $\pm$ 1.4 & 22.7 $\pm$ 5.7 & 14, May 2017 - 2, July 2022 \\
    HM Sge & 54.5 $\pm$ 1.2 & 60.0 $\pm$ 1.4 & 72.8 $\pm$ 5.6 & 14, May 2017 - 2, July 2022 \\
    RR Tel$^*$ & 66.8 $\pm$ 1.0 & 76.8 $\pm$ 1.2 & 73.0 $\pm$ 4.6 & 11, May 2017 - 1, July 2022 \\
    IRAS 20124+1154 & 6.3 $\pm$ 1.3 & 7.9 $\pm$ 1.5 & 6.5 $\pm$ 5.9 & 14, May 2017 - 2, July 2022 \\
    DD Mic$^*$ & 4.1 $\pm$ 1.2 & 5.0 $\pm$ 1.4 & 9.4 $\pm$ 5.7 & 11, May 2017 - 1, July 2022 \\
    AG Peg & 26.9 $\pm$ 1.2 & 29.8 $\pm$ 1.5 & 34.2 $\pm$ 6.1 & 14, May 2017 - 30, June 2022 \\
    R Aqr & 46.7 $\pm$ 1.5 & 52.3 $\pm$ 1.7 & 73.2 $\pm$ 6.5 & 14, May 2017 - 30, June 2022 \\
    \bottomrule
    \end{tabular*}
    \begin{minipage}{\textwidth}
        \raggedright
        $^*$: also seen in SPT. DD Mic and ShWi 5 did not meet detection criteria in ACT, but we include their ACT data due to detections in SPT.
    \end{minipage}
\end{table*}

Cutting NODSV to the nominal footprints of our maps left a total of 828 candidate SySts: 216 in SPT, 400 in ACT, and 212 in the footprint of both surveys. 

Though the passbands differ slightly, we will refer to the SPT and ACT data using the ``95", ``150", and ``220" GHz labels throughout this paper. Treating each dataset independently, we set a detection criteria of $>3\sigma$ measurements in any two of the 95, 150, and 220~GHz bands and then assessed the data and made additional cuts.  First, we removed extragalactic SySts (one in SPT) as there was no reasonable way to discern the detection as the individual SySt or NGC 55 itself at $\sim$2 Mpc.  Second, detections that met the SNR criteria but did not appear to be point sources\footnote{Considering that expanding dust shells/nebulae may result in diffuse emission at mm wavelengths, we note that ACT observations of R Aqr (a nearby SySt at $\sim$280 pc with a resolved nebula) still appeared as a point source and supports this criteria.} or were not centered at the Syst location (two in SPT, 44 in ACT) were removed.  Finally, we removed one source detected in ACT that appears to have been incorrectly associated with a different star in NODSV, which is not a SySt. 

This resulted in 15 out of 428 detections in SPT (Table~\ref{tbl:systs_spt}) and 23 out of 612 detections in ACT (Table~\ref{tbl:systs_act}), including seven detected SySts in both datasets, leading to a total of 31 out of 828 unique SySts detected. Two stars, DD~Mic and ShWi~5, did not pass the detection threshold in ACT but were detected by SPT and are within the ACT footprint so we include their ACT light curve data throughout this analysis. Table \ref{tbl:systs_main} shows the combined SPT+ACT detected SySts along with basic information. For distances, we provide the value along with method used and reference.

\begin{table*}[ht!]
\footnotesize
    \centering
    \caption{\raggedright\label{tbl:systs_main}Information for SySts detected in SPT and ACT, provided from NODSV unless otherwise noted. ``?'' in IR type signifies it is not a confirmed SySt. Distances are a mix of values provided from NODSV as well as our own literature search with methods and references for each measurement given. SySts are separated into groups by their IR types, with all suspected SySts in one group.}
    \begin{tabular*}{\textwidth}{l@{\extracolsep{\fill}}rrcccr}
    \toprule
    \toprule
    Name & R.A.  & Decl. & IR type &  Outbursts & Distance & Distance method\\
          & [deg] & [deg] &        &            & [pc]      &   \\ 
    \midrule
    omi Cet  &  34.8366  &  $-$2.9776 &  D   & - &  110 $\pm$ 18    & Mira period-luminosity relation$^{1}$\\
    V852 Cen  &  212.9669  &  $-$51.4401 &  D   &  -  &  3300 $\pm$ 900    & Nebula expansion parallax$^{2}$\\
    K 5-33  &  266.1248  &  $-$27.3446 &  D    & - &  7087 $\pm$ 3336    & Geometric$^{3}$\\
    H 1-36  &  267.4509  &  $-$37.0244 &  D    & - &  5050 $\pm$ 890    & pAGB luminosity, SED modeling$^{4}$\\
    H 1-45  &  269.5911  &  $-$28.2478 &  D    & - &  6200 $\pm$ 1400    & Mira period-luminosity relation$^{5}$\\
    H 2-38  &  271.5048  &  $-$28.2845 &  D    & - &  7200 $\pm$ 1152   & Mira period-luminosity relation$^{1}$\\
    Hen 2-375  &  274.5376  &  $-$57.1871 &  D    & - &  5634 $\pm$ 2097   & Geometric$^{3}$\\
    K 3-9  &  280.1008  &  $-$8.7295 &  D    & - &  2650 $\pm$ 470   & Geometric$^{3}$\\
    HM Sge  &  295.4878  &  16.7444 &  D    & SyN  &  1036 $\pm$ 110   & Geometric$^{3}$\\
    RR Tel  &  301.0773  &  $-$55.7259 &  D    & SyN  &  2700 $\pm$ 300   & Mira period-luminosity relation$^{6}$ \\
    AS 201  &  127.9287  &  $-$27.7588 &  D'    & - &  4300 $\pm$ 2050   & Spectroscopic luminosity estimate$^{7}$\\
    ShWi 5  &  270.9737  &  $-$29.8562 &  D'    & - &  7042 $\pm$ 1821   & Geometric$^{3}$\\ \\
    CN Cha  &  164.99  &  $-$79.9503 &  S    & SyN &  3051 $\pm$ 184    & Geometric$^{3}$\\
    V455 Sco  &  256.8406  &  $-$34.0874 & S   & Yes$^{*}$ &  2800 $\pm$ 700   & Spectroscopic parallax$^{8}$\\
    AR Pav  &  275.1162  &  $-$66.0786 &  S   & Z~And &  4800 $\pm$ 1000   & Spectroscopic parallax$^{9}$\\
    DD Mic  &  315.0264  &  $-$42.6458 &  S   & No$^{\dagger}$ &  2100 $\pm$ -$^\ddagger$   & Spectroscopic luminosity estimate$^{10}$\\
    AG Peg  &  327.7582  &  12.6256 &  S   & SyN, Z~And &  1272 $\pm$ 50   & Geometric$^{3}$\\ 
    R Aqr  &  355.9562  &  $-$15.2846 &  S   & Yes$^\circ$ &  280 $\pm$ 15   & SiO maser parallax$^{11}$\\ \\
    ASASSN-17dm  &  151.7397  &  $-$47.7283 &  D?  & Z~And? &  3791 $\pm$ 264   & Geometric$^{3}$\\
    IRAS 18344-0632 & 279.2717 &  $-$6.4939 & D? & - & 12600.0 $\pm$ 700.0 & Kinematic$^{12}$ \\
    PM 1-253  &  280.9025  &  3.7778 &  D?   & - &  6214 $\pm$ 3078   & Geometric$^{3}$\\
    Hu 2-1  &  282.4482  &  20.8442 &  D?   & -  &  2373 $\pm$ 342   & Geometric$^{3}$\\
    WISE J192140.40+155354.6  &  290.4184  &  15.8985 &  D? &  - &  -$^\ddagger$   & -\\
    QX Pup  &  115.5695  &  $-$14.714 &  D'? & -  &  1538 $\pm$ 24   & H$_{\mathrm{2}}$O maser parallax$^{13}$\\
    Sa 2-18  &  118.2406  &  $-$36.7318 &  D'?   & - &  -$^\ddagger$   & -\\
    WSTB 19W032  &  264.7621  &  $-$28.9432 &  D'?   & - &  2275 $\pm$ 155   & Geometric$^{3}$\\
    MaC 1-10  &  272.3017  &  $-$25.0763 &  D'?   & - &  3144 $\pm$ 1091   & Geometric$^{3}$\\
    M 1-57  &  280.0844  &  $-$10.6631 &  D'?  & - &  4273 $\pm$ 1300   & Geometric$^{3}$\\
    PM 1-286  &  287.8993  &  13.5198 &  D'? &  - &  8615 $\pm$ 3197   & Geometric$^{3}$\\
    IRAS 20124+1154  &  303.715  &  12.0648&  D'?   & Yes? &  1969 $\pm$ 284    & Geometric$^{3}$\\
    PN Me 1-1  &  294.7909  &  15.9467 &  S? & - &  3639 $\pm$ 254   & Geometric$^{3}$\\
    \bottomrule
    \end{tabular*}
    \begin{minipage}{\textwidth}
        {\raggedright
        \textbf{$^\circ$}: R Aqr has shown historical evidence of multiple bright outbursts dating back millennia, but with no confirmation of outburst type. \\
        \textbf{$^*$}: V455 Sco shows an outburst in the mid 1930s \citep{swope40} but the data are sparse and no evidence is seen in recent data displayed in Figure~\ref{fig:s_lc}. \\ 
        \textbf{$^\dagger$}: DD Mic shows periodic brightenings but as described in Section~\ref{sec:confirmed_s}, this relates to the orbital period and not thermonuclear ignition.\\
        \textbf{$^\ddagger$}: DD Mic has no uncertainty on distance provided, while WISE J192140.40+155354.6 and Sa 2-18 have no distance estimates available in the literature. \\
        {\footnotesize 1:\citet{gromadzki09}, 2:\citet{santandergarcia08}, 3:\citet{bailerjones21}, 4:\citet{vickers14}, 5:\citet{miszalski13}, 6:\citet{jurkic12}, 7:\citet{pereira05}, 8:\citet{mikolajewska97}, 9:\citet{fekel17}, 10:\citet{skopal05}, 11:\citet{min14}, 12: \citet{anderson15} 13:\citet{choi12}}}
        \end{minipage}
\end{table*}

We note that there are probable real detections below our threshold criteria, as multiple factors (filtering methods, locations near map edge, etc.) can affect this. In particular, the noise estimates in SPT can be influenced by overdensities of nearby bright sources included in the annulus leading to an inaccurate RMS of local noise. For this reason, we provide the thumbnails and light curve data for all SySts in the SPT and ACT fields, including non-detections at \url{https://doi.org/10.13012/B2IDB-4160106_V1}. We also make brief comments on the non-detections of confirmed D- and D'-type SySts in Section~\ref{sec:appendix}.

 \begin{figure*}[ht]
    \centering
    \includegraphics[width=\textwidth,center]{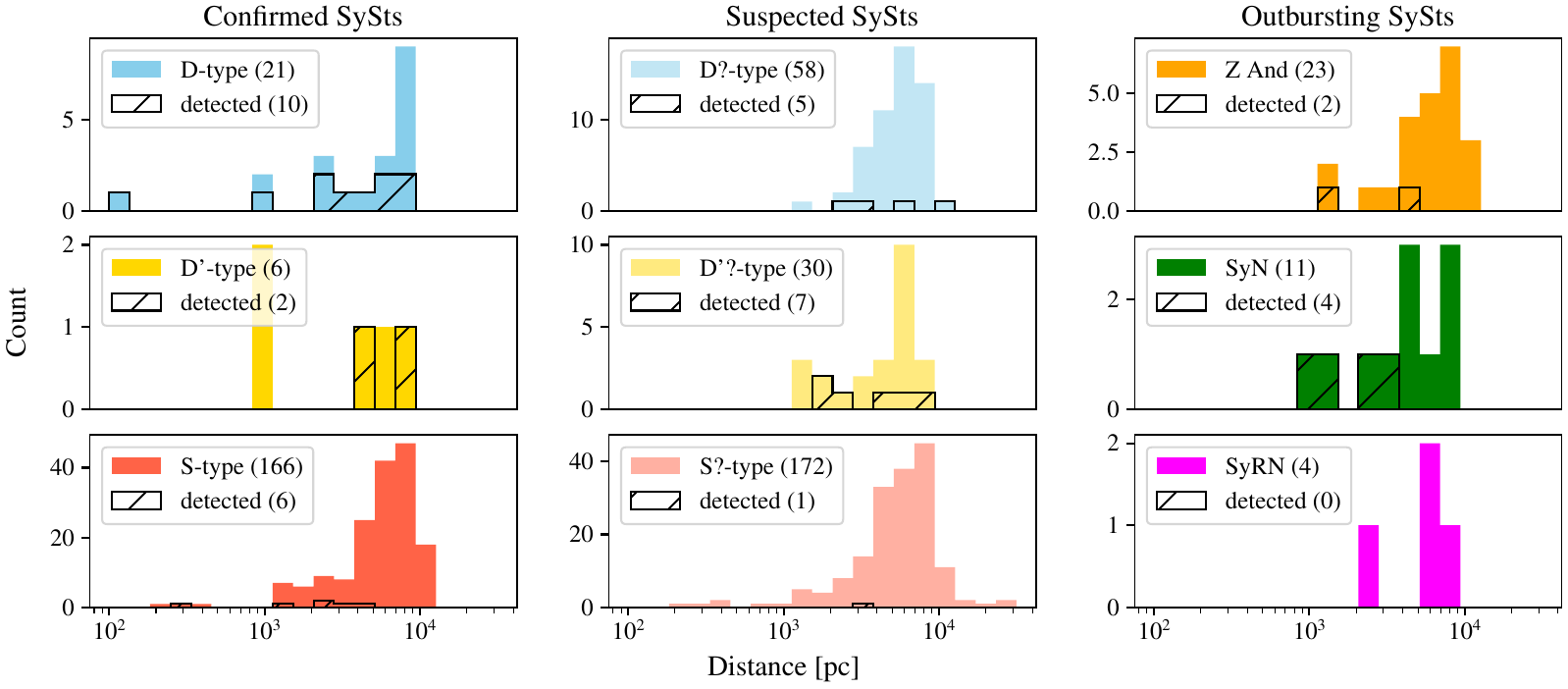}
    \caption{Distance histograms comparing detected vs non-detected MW SySts in various categories.}
    \label{fig:detections}
\end{figure*}

Figure~\ref{fig:detections} shows histograms comparing the SPT/ACT detected SySts to non-detections, binned by distance, and split into the categories of IR type for confirmed and suspected SySts, as well as SySts showing evidence of outbursts. We exclude extragalactic SySts from the total populations. For IR type we exclude the following: 312 SySts with no designation, 4 as ``D/S'', 2 as ``S,D'', and 1 as ``S/D?''. For outbursting SySts we exclude any uncertain outburst types (both in NODSV, and in Table~\ref{tbl:systs_main}), while also including types classified as ``SyN + Z~And'' and ``SyN or Z~And'' into both the SyN and Z~And histograms. Therefore the total numbers do not add up to the 828 candidate SySts within the SPT/ACT footprint.

\begin{figure}[ht!]
    \centering
    \includegraphics[width=\columnwidth,center]{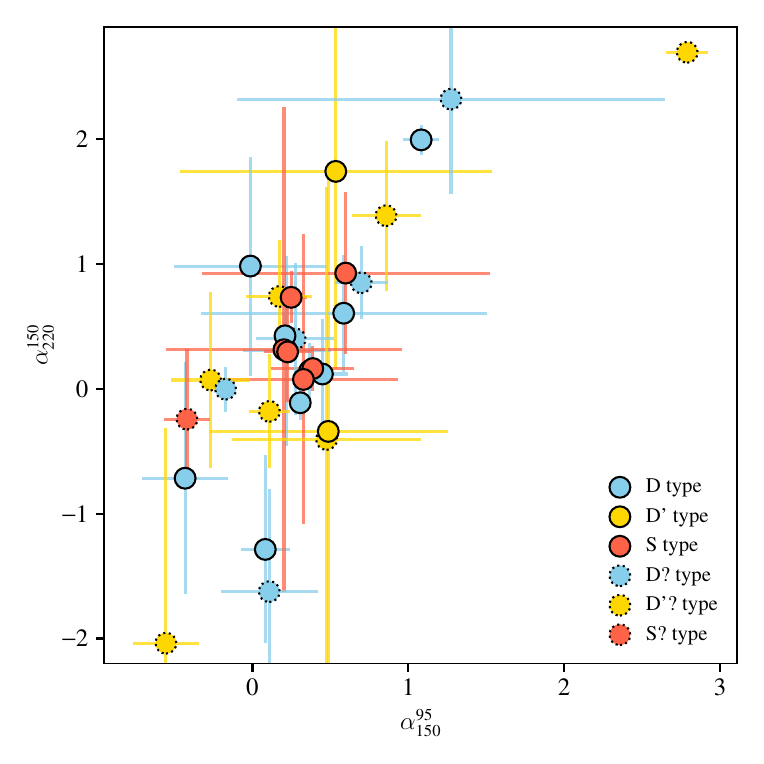}
    \caption{mm spectral indices of SySts between the 95/150~GHz~And 150/220~GHz bands. SySts detected by both SPT and ACT use spectral indices from the higher SNR detection.}
    \label{fig:spec_idx}
\end{figure}

We overwhelmingly detect confirmed D- (47.6\%) and D'-type (33.3\%) SySts at a higher rate than S-types (3.6\%), with suspected SySts for each IR type being detected at a lower rate than their confirmed counterparts. Suspected SySts follow a similar trend: D? (8.6\%), D'? (23.3\%), and S? (0.6\%). The lower rate of detections compared to their confirmed counterparts could be related to selection effects. The most well-studied systems are generally either the ones which have shown prominent novae, or the ones which are nearest and brightest.  Furthermore, if there is substantial contamination in the unconfirmed sample, and the contaminants like PNe or H\,{\sc ii} regions tend to be weaker at mm wavelengths than symbiotics of similar brightness in the band used for selection, this, too would lead to lower detection rates. 

Because the effects of extinction at mm wavelengths are negligible, the effects of distance come in only via the inverse-square law.  Even this effect is not apparent in the observed samples, which may be indicative of the range of luminosities being larger than the range of values of $d^2$ in the current samples.  The six S-type SySts detected in SPT and ACT are known outbursting systems (SyN in CN Cha, Z~And in AR~Pav, SyN + Z~And in AG~Peg), have outbursts without classification (V455~Sco), have shown strong variability (periastron passage, DD~Mic), or have plausible historical evidence of outbursts (R Aqr); we individually remark on the nature of these further in Section~\ref{sec:confirmed_s}.

Transient analysis of the SPT-3G Galactic Plane Survey resulted in the detection of flaring events in two accreting white dwarf systems, one of which is a SySt \citep{wan26}. The SySt (2SXPS~J173508.4-292958 = Gaia~DR3~4058581921170339456) was observed flaring over the duration of $\sim1$ day with peak flux densities of $\sim60$ and $\sim90$ mJy in the 95 and 150 GHz bands, respectively. However it was not detected in this work, as the coadded SNR was well below the threshold in all three bands and no obvious point source was visible in the thumbnails.

Figure~\ref{fig:spec_idx} shows the 2-band mm spectral indices of the SPT/ACT detected SySts. mm/sub-mm observations can provide a model-independent method of determining $\nu_t$, but require careful untangling of the contributing mechanisms which is beyond the scope of this paper. As mentioned previously, D- and D'-type SySts are primarily a combination of optically thin FF ($\alpha=-0.1$) and blackbody emission ($\alpha$=2); the SPT/ACT detected population is in general agreement with this. For the six S-type SySts we detected --- CN Cha, V455 Sco, AR Pav, DD Mic, AG Peg, and R Aqr --- SEDs (Figure~\ref{fig:confirmed_sed}) show rising spectra at mm wavelengths that cannot be explained by the thermal blackbody emission of the giant's photosphere alone. We interpret these spectral indices as evidence for a significant contribution from optically thick FF emission, though we stress that all six detected S-type SySts have confirmed outbursts (or other strong variability features that have been confused for outbursts) and may not agree with the STB model as they may not meet the standard definition of ``quiescent.''

Figure~\ref{fig:lum_since_outburst} shows the mm luminosities of SySts exhibiting SyN and Z~And outbursts as a function of time since their outbursts. DD Mic is excluded since its variability is related to periastron passage which is explained further in Section~\ref{sec:confirmed_s}. All light curves have data in one-year bins with the exceptions of AR Pav, ASASSN-17dm, and CN Cha, which are binned to 14 days for visual clarity. The type of outburst precedes the names of the SySt, with question marks indicating unconfirmed SySts (IRAS 20124+1154, ASASSN-17dm) or uncertainty on type of outburst (V455 Sco). As mentioned in Section \ref{sec:external_data}, geometric distance may be unreliable due to uncertainties in parallax estimates; IRAS 20124+1154, ASASSN-17dm, AG Peg, CN Cha, and HM Sge use geometric distance to calculate their mm luminosities in Figure \ref{fig:lum_since_outburst}.  Of those objects, CN Cha and AG~Peg are S-type systems, so they are most susceptible to astrometric wobble problems.

Table \ref{tbl:nova_dates} shows the dates used as onset of outburst for Figure \ref{fig:lum_since_outburst}. For AG~Peg, we use the 2015 Z~And outburst \citep{tomov16} rather than the $\sim$1850 SyN \citep{lundmark12} as the most recent outburst; the ACT data shows a prominent decay in all three bands which is likely resultant from the recent Z~And outburst rather than the SyN (see Section \ref{sec:confirmed_s} for more details). HM~Sge shows a rise in 220~GHz luminosity which may indicate dust formation, and is discussed further in Section \ref{sec:confirmed_d}. RR~Tel continues to show mild decay $\sim$80 years after its nova (Section \ref{sec:confirmed_d}), with SPT-SZ data from 2009 giving a clear reference flux in all three bands nearly 10 years prior to ACT data. R Aqr does not have a confirmed nova on this date, but measurements of the nebula suggest the origins are from repeated outbursts with the most recent happening in the early 19th century (Section \ref{sec:confirmed_d}.)

\subsection{CN Cha: a potential rare Galactic mm slow transient}

\begin{figure}[ht!]
    \centering
    \includegraphics[width=\columnwidth,center]{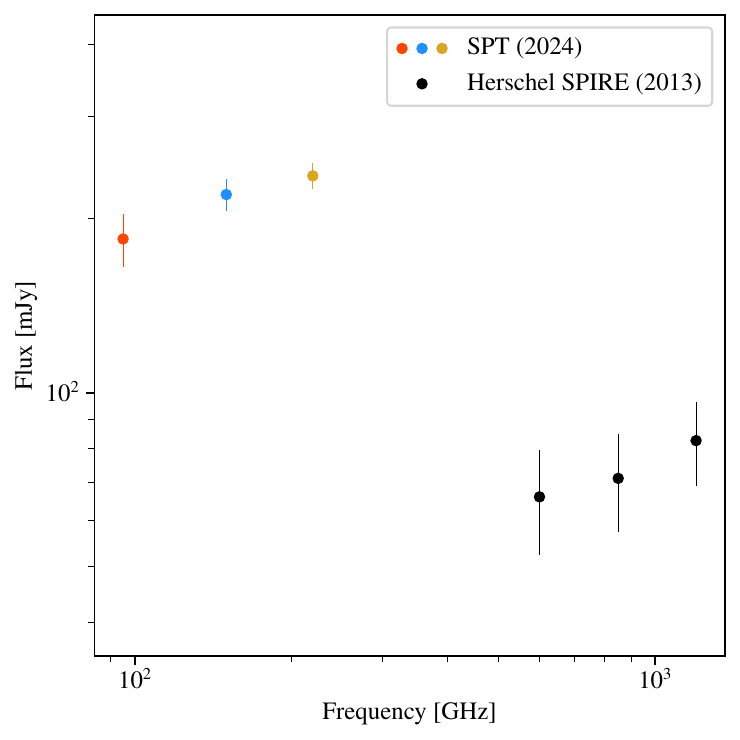}
    \caption{Forced photometry measurements of CN Cha from Herschel SPIRE (2013) and SPT (2024), suggesting strong mm variability occurred around the time of the optical nova in 2012/2013.}
    \label{fig:cn_cha_spt_herschel}
\end{figure}

The field of CN~Cha was also observed with Herschel SPIRE \citep{griffin10} in March of 2013. CN~Cha was not reported as a detection in the source catalogs,\footnote{\url{https://doi.org/10.5270/esa-6gfkpzh}} but was covered by the detectors.  The background in this region of the sky shows some structure, making it hard to estimate the source flux density precisely. Forced photometry at the location of CN Cha in the three Herschel bands of 600, 850, and 1200~GHz thus gives us upper limits of $66\pm14$, $71\pm14$, and $83\pm14$ mJy, respectively.\footnote{Data used are level 2 products, with Observation ID 1342265327 and Target Name \texttt{PLCK\_SY\_G298.0-18.3}.} Noise was calculated taking an RMS of an annulus around the location, in each band. SPT flux values in 2024 are $184\pm19$, $220\pm14$, and $237\pm13$ mJy at 95, 150, and 220~GHz, respectively. The spectral index between 220~GHz in 2024 and 600 GHz in 2013 would be $-1.3\pm0.2$ if the source were non-variable. Figure \ref{fig:cn_cha_spt_herschel} shows these flux measurements.

This would not be possible for FF emission---for gas to be ionized enough to produce FF emission, it must be hot enough to have a cutoff in the near-infrared or higher frequency bands. Furthermore, in the Herschel bands, it is likely that the thermal tail of the AGB star's dust reprocessed emission is starting to contribute, so the spectrum would have to be steeper, indicative of cooled synchrotron.

\begin{table}
\centering
\footnotesize
\caption{\raggedright\label{tbl:nova_dates}Dates used as onset of outburst for Figure \ref{fig:lum_since_outburst}. If exact dates weren't supplied, a reasonable approximation was chosen.}
\begin{tabular*}{\columnwidth}{l@{\extracolsep{\fill}}rr}
    \toprule
    \toprule
    Name            &   Date    &   Source  \\ 
    \midrule
    IRAS 20124+1154 &  25 Feb, 2021 &   \citet{paunzen23}       \\
    AG Peg          &   1 Jun, 2015 &   \citet{tomov16}         \\
    AR Pav          &   1 Jun, 2021 &   \citet{merc21}          \\
    ASASSN-17dm     &  21 Feb, 2017 &   \citet{kiyota17}        \\   
    CN Cha          &  31 Dec, 2012 &   \citet{lancaster2020}   \\
    HM Sge          &   1 Sep, 1975 &   \citet{dokuchaeva76}    \\
    RR Tel          &   1 Oct, 1944 &   \citet{mayall49}        \\      
    V455 Sco        &  16 Jul, 1935 &   \citet{swope40}         \\
    R Aqr           &   1 Jan, 1802 &   \citet{solf85}          \\       
    \bottomrule
\end{tabular*}
\end{table}

As a result of this, it seems apparent that the weakness of the Herschel emission implies that the emission in the SPT bands must also have been well below current levels in March of 2013, if indeed the source were variable
The optical nova had already started at the time of the Herschel observations.  Thus, the Herschel data indicate both that the source may have varied strongly at mm wavelengths, and, if so, that this rise must have taken place with a lag with respect to the optical rise.

CN Cha would be just the second Galactic mm slow transient to be observed from an untriggered search. The classical nova YZ Ret was observed during outburst in ACT data in 2020, about 60 days after the initial X-ray flash at the start of the nova \citep{biermann25}. They also provide the two other mm observations of classical novae---Nova Cygni in 1992 by \citet{ivison93} and ALMA observations of the remnant of Nova V5668 Cyg, 3 years after its nova in 2015 by \citet{diaz18}---highlighting the rarity of mm observations of such events.


\section{Conclusion} \label{sec:conclusion}
In this paper we have presented the results of forced photometry on the location of 828 candidate SySts in SPT-3G and ACT~DR6 data, with a total of 31 unique systems in the MW detected between the two datasets. We have found that FF emission, both optically thick and thin, as well as optically thick blackbody emission contribute significantly to the flux levels at 95, 150, and 220~GHz. FF emission and the turnover frequency, $\nu_t$, as part of the STB binary model have been shown to be crucially important to understanding the geometry of the ionizing region between the binary components. 

While $\nu_t$ is typically $\sim$1-10~GHz for D- and D'-type SySts, the higher frequencies of SPT and ACT can still play a strong role in disentangling contributions from optically thin FF and optically thick blackbody emission in these stars. S-types are found to have higher $\nu_t$ on the order of 10$^3$ GHz due to the shorter binary separation, but are under-studied at mm wavelengths and can offer insights into the behavior of gas and dust in the system, along with mass loss of the RG overall. Investigations into the Galactic Plane with radio and mm surveys can probe the high extinction regions where optical spectroscopy cannot. For PNe and D/D'-type SySts in these regions, constraining $\nu_t$ can result in measurements of radius and EM of their nebulae. This could allow for an additional diagnostic metric, and could potentially result in new SySts being discovered that would not be possible using typical shorter wavelength observations.

Our detections are in agreement with historical studies at similar wavelengths, finding that the mm regime is more sensitive to D- and D'- type systems than S-types due to the excess dust present, despite the fact that S-types in the MW have been classified in SySt catalogs at much higher rates. We emphasize that this discrepancy in the lower number of D- and D'-type SySts compared to S-type SySts are likely influenced by selection effects and biases such as intrinsic dust reddening in these systems when using optical and IR instruments for detection.

CN Cha experienced a Symbiotic Nova in late 2012/early 2013, and was observed by Herschel SPIRE in March, 2013. We have shown the data from SPIRE strongly suggests a significant increase in flux between 2013 and 2024 as well as a lag compared to the optical rise. If the source were variable, this would be the second Galactic mm slow transient detection from an untriggered search after the classical nova YZ Ret in \citet{biermann25}.

We provide multi-wavelength light curves and SEDs along with a brief description of each detected SySt. As different filtering methods and noise estimations can affect the SNR used in our detection threshold, we also include the SPT/ACT light curves and thumbnails for all 828 candidate SySts at \url{https://doi.org/10.13012/B2IDB-4160106_V1}.

\begin{figure*}[t]
    \centering
    \includegraphics[width=\textwidth,center]{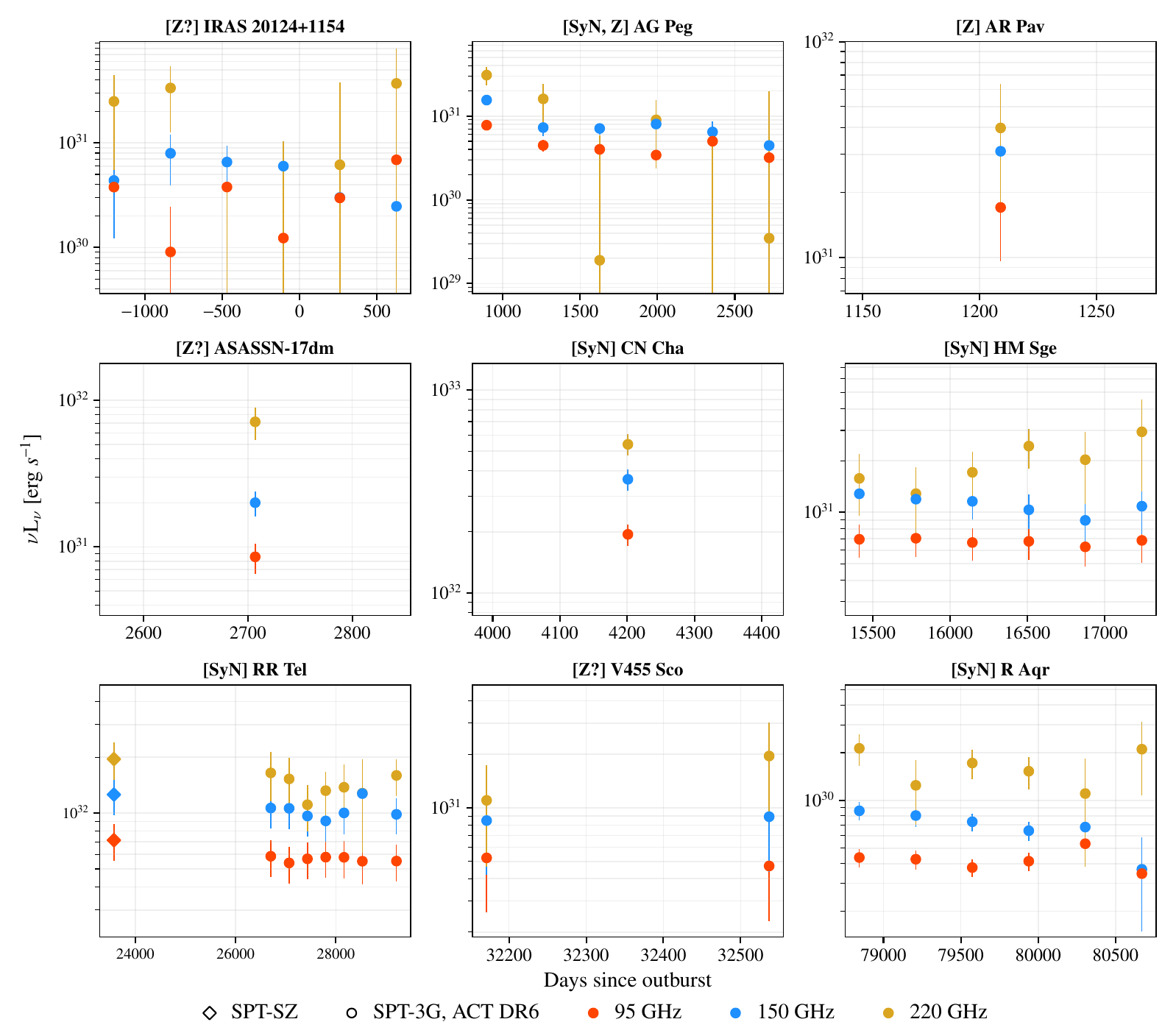}
    \caption{mm luminosities of each outbursting SySt as a function of time since their most recent outburst, with associated outburst type(s) included next to SySt names. The light curve for RR Tel includes SPT-SZ data from 2009 as described in Section \ref{sec:confirmed_d}. Luminosities are binned by year to emphasize any potential decaying rather than to highlight variability. IRAS 20124+1154, ASASSN-17dm, AG Peg, CN CHa, and HM Sge use geometric distance to calculate their mm luminosities, which may contain additional uncertainties as described in Section \ref{sec:external_data}.}
    \label{fig:lum_since_outburst}
\end{figure*}

The work in this paper represents the latest example of Galactic astronomy made possible with CMB surveys. These data span nearly a decade of mm/sub-mm observations in the southern hemisphere (2019--2025 in SPT and 2017--2022 in ACT). SPT and ACT can provide data for MW objects, either to be used to strengthen proposals for targeted observations, or as an easier alternative to scheduling time on observatories such as ALMA. Future CMB surveys like the Simons Observatory (SO) have forecasted higher detection rates of Galactic mm transients \citep{abitbol25} and have planned studies of the Galactic Plane \citep{hensley22}; additionally, extra observing bands in SO---with nominal values of 27, 39, and 280~GHz---could lead to better understanding of the mm/sub-mm SED. As shown in Figure~\ref{fig:skyplot}, candidate SySts are primarily located in the Galactic Plane, and specifically the Galactic Center, i.e. regions of extremely high stellar density. Future upgrades to the SPT camera with more detector plane solid angle---leading to larger effective exposure times---could lead to better detections in these crowded regions than SO would be capable of due to the worse angular resolution. Extragalactic SySts are potentially different in nature to those in the MW, and are understudied. As new methods of detecting them (e.g. narrow-band O\,{\sc vi} filter upgrades to already-existing telescopes) are explored, mm/sub-mm observations could provide valuable information. An upgraded SPT could also study the Magellanic Clouds, which are too far south for SO to observe, and would provide complementary data to SO in the coming decade. 


\section*{Acknowledgments}\label{sec:acknowledgments}
We thank the anonymous reviewer for their thorough reading of the manuscript and helpful comments. We thank Kedron Sillsbee for useful discussions about spinning dust and Elias Aydi and Kirill Sokolovsky for useful discussions about symbiotic novae and Jaroslav Merc for useful communications regarding his catalog.  We also thank Anthony Gonzalez and Ariane Trudeau for useful discussion about Herschel systematics, as well as Joanna Miko{\l}ajewska, Lev Yungelson, and Rob Ivison for valuable feedback during the submission process.

The South Pole Telescope program is supported by the National Science Foundation (NSF) through awards OPP-1852617 and OPP-2332483. Partial support is also provided by the Kavli Institute of Cosmological Physics at the University of Chicago. 

Argonne National Laboratory’s work was supported by the U.S. Department of Energy, Office of High Energy Physics, under contract DE-AC02-06CH11357. 

The UC Davis group acknowledges support from Michael and Ester Vaida. 

Work at the Fermi National Accelerator Laboratory (Fermilab), a U.S. Department of Energy, Office of Science, Office of High Energy Physics HEP User Facility, is managed by Fermi Forward Discovery Group, LLC, acting under Contract No. 89243024CSC000002.

The Melbourne authors acknowledge support from the Australian Research Council’s Discovery Project scheme (No. DP210102386). 

The Paris group has received funding from the European Research Council (ERC) under the European Union’s Horizon 2020 research and innovation program (grant agreement No 101001897), and funding from the Centre National d’Etudes Spatiales. 

The SLAC group is supported in part by the Department of Energy at SLAC National Accelerator Laboratory, under contract DE-AC02-76SF00515.

We gratefully acknowledge the contributions of the AAVSO observer community, whose photometric data and metadata resources were used in this study and made available through the AAVSO’s scientific archives.

This work has made use of data from the Asteroid Terrestrial-impact Last Alert System (ATLAS) project. The Asteroid Terrestrial-impact Last Alert System (ATLAS) project is primarily funded to search for near earth asteroids through NASA grants NN12AR55G, 80NSSC18K0284, and 80NSSC18K1575; byproducts of the NEO search include images and catalogs from the survey area. This work was partially funded by Kepler/K2 grant J1944/80NSSC19K0112 and HST GO-15889, and STFC grants ST/T000198/1 and ST/S006109/1. The ATLAS science products have been made possible through the contributions of the University of Hawaii Institute for Astronomy, the Queen’s University Belfast, the Space Telescope Science Institute, the South African Astronomical Observatory, and The Millennium Institute of Astrophysics (MAS), Chile.

This research has made use of the VizieR catalogue access tool, CDS, Strasbourg, France \citep{10.26093/cds/vizier}. The original description of the VizieR service was published in \citet{vizier2000}.

This research has made use of the NASA/IPAC Infrared Science Archive, which is funded by the National Aeronautics and Space Administration and operated by the California Institute of Technology. This publication makes use of data products from the Wide-field Infrared Survey Explorer, which is a joint project of the University of California, Los Angeles, and the Jet Propulsion Laboratory/California Institute of Technology, and NEOWISE, which is a project of the Jet Propulsion Laboratory/California Institute of Technology. WISE and NEOWISE are funded by the National Aeronautics and Space Administration. 

Herschel is an ESA space observatory with science instruments provided by European-led Principal Investigator consortia and with important participation from NASA.

This research was done using services provided by the OSG Consortium \citep{osg07, osg09, osg06, osg15}, which is supported by the National Science Foundation awards \#2030508 and \#2323298. 
This work made use of the following software packages: \texttt{astropy} \citep{astropy:2013,astropy:2018,astropy:2022}, \texttt{Jupyter} \citep{2007CSE.....9c..21P,kluyver2016jupyter}, \texttt{matplotlib} \citep{Hunter:2007}, \texttt{numpy} \citep{numpy}, \texttt{pandas} \citep{mckinney-proc-scipy-2010,pandas_10957263}, \texttt{python} \citep{python}, \texttt{scipy} \citep{2020SciPy-NMeth,scipy_10909890}, and \texttt{astroquery} \citep{2019AJ....157...98G,astroquery_10799414}.
This research has made use of the Astrophysics Data System, funded by NASA under Cooperative Agreement 80NSSC21M00561.
Software citation information aggregated using \texttt{\href{https://www.tomwagg.com/software-citation-station/}{The Software Citation Station}} \citep{software-citation-station-paper,software-citation-station-zenodo}.


\bibliographystyle{yahapj}
\bibliography{refs}


\appendix 
\section{} \label{sec:appendix} 
\setcounter{figure}{0}
\renewcommand{\thefigure}{A\arabic{figure}}
\makeatletter
\renewcommand{\theHfigure}{A\arabic{figure}}
\makeatother
\setcounter{table}{0}
\renewcommand{\thetable}{A\arabic{table}}
\makeatletter
\renewcommand{\theHtable}{A\arabic{table}}
\makeatother

Table~\ref{tbl:non_detections} shows confirmed MW D- and D'-type SySts that did not pass the detection thresholds in SPT and ACT. Comments result from visual inspection of thumbnails: ``Near threshold'' means a point source appears in the center of the image, but falls short of the 2-band $3\sigma$ significance threshold, with ``ambiguous'' indicating there is not a confident claim of said point source. ``No source'' indicates there is neither a point source nor did the location come close to the detection threshold. As S-type SySts are considerably more numerous than D- and D'-type (Figure~\ref{fig:detections}), we exclude them from this table along with all suspected D- and D'-type SySts. We include the Galactic coordinates to show that most non-detections occur in the crowded Galactic Plane, where accurate noise estimation is difficult. 

\begin{table*}[h!]
    \footnotesize
    \centering
    \caption{\raggedright\label{tbl:non_detections}Non-detections of confirmed D- and D'-type MW SySts. SySts have been separated into groups based on IR type, similar to Table \ref{tbl:systs_main}.}
    \begin{tabular*}{\textwidth}{l@{\extracolsep{\fill}}lllllll}
    \toprule
    \toprule
    Star name  & IR type &  b    & l     & Comment & Telescope \\
            &            & [deg] & [deg] &         &           \\
    \midrule
    MaC 1-3                 &   D  &    339.6188  &    $-$3.5189 &  No source  &  SPT  \\
    Hen 2-251               &   D  &    358.1205  &       1.4588 &  No source  &  ACT, SPT  \\
    JaSt 2-6                &   D  &    359.9659  &    $-$1.1446 &  No source  &  ACT, SPT  \\
    WRAY 16-312             &   D  &    358.7885  &    $-$1.9092 &  No source  &  ACT, SPT  \\
    JaSt 79                 &   D  &      0.2093  &    $-$1.4710 &  Near threshold  &  ACT, SPT  \\
    V5590 Sgr               &   D  &      4.2275  &    $-$4.0402 &  Near threshold  &  ACT, SPT  \\
    SRGA J181414.6-225604   &   D  &     8.4036   &    $-$2.5960 &  No source$^X$  &  ACT  \\
    V3929 Sgr               &   D  &      5.6988  &    $-$5.7617 &  Near threshold, ambiguous &  ACT  \\
    Sct X-1                 &   D  &     24.3360  &       0.0657 &  No source$^X$  &  ACT  \\
    K 3-22                  &   D  &     45.6591  &       1.5232 &  Near threshold, ambiguous  & ACT  \\
    EF Aql                  &   D  &     34.6144  &   $-$16.0731 &  No source  &  ACT  \\ \\
    GH Gem                  &  D'  &    203.5630  &       8.2653 &  No source  &  ACT  \\
    WRAY 15-157             &  D'  &    246.6048  &       1.9473 &  No source  &  ACT, SPT  \\
    AS 269                  &  D'  &    358.6650  &    $-$5.1961 &  No source  &  ACT, SPT  \\
    StHA 190                &  D'  &     58.4162  &   $-$35.4321 &  Near threshold  &  ACT  \\
    \bottomrule
    \end{tabular*}
    {\raggedright\textit{$^X$} SRGA J181414.6-225604 and Sct X-1 are SyXB with NS accretors, see text in Section \ref{sec:appendix} for more details.}
\end{table*}

As mentioned previously, some SySts have NS accretors (or even more rarely, black hole accretors); these are all hard X-ray emitters and thus are known as Symbiotic X-ray Binaries (SyXBs) \citep{masetti06}. NS formation is the result of a massive star undergoing core-collapse supernova, and when part of a symbiotic binary, they arrive at the symbiotic designation from a different evolutionary path than those of intermediate mass stars. SyXB have been detected at much lower numbers than SySts, with $\sim$10-20 observed or suspected in the literature \citep{lu12,yungelson19}. Additionally, the SySts which have been detected in the radio in the past are typically either very faint \citep{2018MNRAS.474L..91V}, or are black hole accretors with Roche-lobe filling donors that are at the faint end of the range of red giants, and where the radio emission is likely from the relativistic jet launched by the black hole accretion disk \citep{1992ApJ...400..304H,1994Natur.371...46M}, rather than dust or a hot wind from the donor star. 

\begin{table*}[h!]
    \footnotesize
    \centering
    \caption{\raggedright\label{tbl:mm_studies}Comparison of mm SySt observations between \citet{ivison95b, mikolajewska02}$^*$ and this work. SySts are separated into groups by their IR type. Yes/No refer to detections by the observing telescope(s), while a dash signifies the object was not observed by that telescope. Information for non-detected IRAM SySts was not readily available.}
    \begin{tabular*}{\textwidth}{l@{\extracolsep{\fill}}llllll}
    \toprule
    \toprule
    Star name               &IR type&  JCMT & SEST  & OVRO  & IRAM  & SPT/ACT (this work)   \\
    \midrule
    AX Per                  &   S   &   No  &   -   &   -   &   -   & -        \\
    BX Mon                  &   S   &   No  &   -   &   -   &   -   & No       \\
    V2416 Sgr               &   S   &   No  &   -   &   -   &   -   & No       \\
    AP 1-8                  &   S   &   No  &   -   &   -   &   -   & No   \\
    AS 289 = V343 Ser       &   S   &   Yes &   -   &   -   &   Yes & No  \\
    CH Cyg                  &   S   &   Yes &   -   &   -   &   Yes & -        \\
    R Aqr                   &   S   &   Yes &   Yes &   -   &   Yes &   Yes   \\
    V1329 Cyg               &   S   &   No  &   -   &   No  &   -   & -        \\
    AG Peg                  &   S   &   No  &   -   &   Yes &   Yes & Yes       \\
    Z And                   &   S   &   No  &   -   &   Yes &   Yes & -        \\
    BF Cyg                  &   S   &   -   &   -   &   No  &   Yes & -        \\
    CI Cyg                  &   S   &   -   &   -   &   No  &   Yes & -        \\
    EG And                  &   S   &   -   &   -   &   -   &   Yes & -        \\
    UV Aur                  &   S   &   -   &   -   &   -   &   Yes & -        \\
    AG Dra                  &   S   &   -   &   -   &   -   &   Yes & -        \\
    V443 Her                &   S   &   -   &   -   &   -   &   Yes & -        \\
    RW Hya                  &   S   &   -   &   -   &   -   &   Yes & No       \\
    FN Sgr                  &   S   &   -   &   -   &   -   &   Yes & No       \\
    V2416 Sgr               &   S   &   -   &   -   &   -   &   Yes & No       \\
    CL Sco                  &   S   &   -   &   -   &   -   &   Yes & -        \\
    RT Ser                  &   S   &   -   &   -   &   -   &   Yes & No       \\
    PU Vul                  &   S   &   -   &   -   &   -   &   Yes & -        \\
    HD 319167               &   S   &   -   &   -   &   -   &   Yes & No   \\
    Hen 1341 = V2523 Oph    &   S   &   -   &   -   &   -   &   Yes & No       \\
    Hen 1591 = Hen 3-1591   &   S   &   -   &   -   &   -   &   Yes & No   \\
    M 1-21                  &   S   &   -   &   -   &   -   &   Yes & No       \\ 
    SS 122 $^a$             &   S   &   -   &   -   &   -   &   Yes &   -   \\ \\

    RX Pup                  &   D   &   Yes &   No  &   -   &   -   &   -   \\
    H 1-36                  &   D   &   No  &   -   &   -   &   -   &   Yes   \\
    K 3-9                   &   D   &   Yes &   -   &   -   &   -   &   Yes   \\
    HM Sge                  &   D   &   No  &   -   &   Yes &   Yes &   Yes   \\
    V1016 Cyg               &   D   &   No  &   -   &   Yes &   Yes &   -   \\
    V407 Cyg $^b$           &   D   &   No  &   -   &   -   &   -   &   -   \\
    He 2-38                 &   D   &   -   &   Yes &   -   &   -   &   Yes   \\
    BI Cru                  &   D   &   -   &   Yes &   -   &   -   &   -   \\
    SS38 = SS73 38          &   D   &   -   &   No  &   -   &   -   &   -   \\
    He 2-104 = V852 Cen     &   D   &   -   &   Yes &   -   &   -   &   Yes   \\
    V835 Cen                &   D   &   -   &   Yes &   -   &   -   &   -   \\
    He 2-171 = Hen 2-171    &   D   &   -   &   Yes &   -   &   -   &   -   \\
    He 2-176 = Hen 2-176    &   D   &   -   &   Yes &   -   &   -   &   No   \\
    AS 210 = V1196 Sco      &   D   &   -   &   Yes &   -   &   -   &   -   \\
    RR Tel                  &   D   &   -   &   Yes &   -   &   -   &   Yes   \\
    H 2-38                  &   D   &   -   &   -   &   Yes &   -   &   -   \\
    K 3-9                   &   D   &   -   &   -   &   Yes &   Yes &   Yes   \\ \\

    Wra 157 = WRAY 15-157   &   D'  &   No  &   -   &   -   &   -   &   No   \\
    AS 201                  &   D'  &   No  &   No  &   -   &   -   &   Yes   \\
    V741 Per = V471 Per     &   D'  &   No  &   -   &   -   &   Yes &   -   \\ \\
    
    HD 149427 $^c$           &   PN  &   -   &   No  &   -   &   -   &   -   \\
    \bottomrule
    \end{tabular*}
    {\raggedright
    \textit{$^*$}: Additional information from \citet{mikolajewska98} and private communication. \\
    \textit{$^a$}: SS 122 is not in NODSV and was not searched for in SPT/ACT data. \\
    \textit{$^b$}: V407 Cyg is classified as an S?-type in \citet{ivison95b} but is classified as a D-type in NODSV. \\
    \textit{$^c$}: HD 149427 is misclassified and is a PN in NODSV.\\
    }
\end{table*}

Table \ref{tbl:mm_studies} shows a comparison of this work to previous investigations of SySts at mm wavelengths. \citet{ivison95b} observed 34 SySts with the James Clerk Maxwell Telescope (JCMT) at 231/273/375/666 GHz, Owens Valley Radio Observatory (OVRO) at 88 GHz, and the Swedish-ESO Submillimetre Telescope (SEST) at 231 GHz. They also include observations using the Australia Telescope Compact Array (ATCA) and the Very Large Array (VLA), however we exclude them as they observe at cm wavelengths with sensitivities that are orders of magnitude lower than SPT/ACT and thus are not suitable for an apt comparison. Comparing the same population of SySts observed between telescopes and this work show similar detection rates between JCMT/OVRO/SEST and this work.

\citet{mikolajewska02} observed 37 SySts with the Institute for Radio Astronomy in the Millimetre Range (IRAM) 30m telescope, and while D- and D-type detection rates (4 out of 5, and 1 out of 2, respectively) are similar to the SPT/ACT detection rates discussed in Section \ref{sec:discussion}, S-type SySts were detected at a much higher rate: 16 out of 30 (53.3\%). Nearly 30 years separate these observations, and variability over those timescales has been shown to potentially have a significant effect. The IRAM observations have a similar uncertainty to SPT/ACT with an rms of $\sim$2~mJy, but do not have the same 2-band $3\sigma$ significance threshold that we have imposed on this work. 

For the case of RW~Hya, an S-type SySt, IRAM observations show a flux value of $\sim$10~mJy. This measurement, and their estimate of optically thick FF emission are roughly in agreement with SPT measured flux values of 2.87/1.96/11.83~mJy at 95/150/220~GHz, respectively, in 2024. While it appears as a point source in SPT thumbnails, it did not pass our SNR threshold having 2.12/0.99/3.58 SNR across the 3 bands. No other candidate SySts in \citet{ivison95b} or \citet{mikolajewska02} were similarly close to being a detection.

\section{Individual Stars} \label{sec:individuals}
\setcounter{figure}{0}
\renewcommand{\thefigure}{B\arabic{figure}}
\makeatletter
\renewcommand{\theHfigure}{B\arabic{figure}}
\makeatother
\setcounter{table}{0}
\renewcommand{\thetable}{B\arabic{table}}
\makeatletter
\renewcommand{\theHtable}{B\arabic{table}}
\makeatother

In this section we provide brief summaries for each star along with any pertinent information from the literature (including other/historical mm data), organized by their IR type and confirmation status. Names of these stars come from NODSV, and where appropriate we use alternate identifiers indicated by an equals sign in the sub-section heading. 

SEDs in Figures \ref{fig:confirmed_sed} and \ref{fig:suspected_sed} are provided for each SySt, with data from a 5\arcsec radius search in VizieR alongside SPT/ACT data from this work.
Figures~\ref{fig:d_lc_1},~\ref{fig:d_lc_2},~\ref{fig:s_lc},~\ref{fig:susp_lc_1}, and ~\ref{fig:susp_lc_2} show multi-wavelength light curves for each SySt from 2016 through 2026, in the following bands: 95, 150, 220~GHz (SPT, ACT), \textit{W1}, \textit{W2} (WISE), \textit{V}, \textit{g} (ASAS-SN) and \textit{o}, \textit{c} (ATLAS). SPT and ACT data have been combined into 3-month bins for visual clarity. When SPT and ACT have observed the same stars, we have not binned any overlapping data between the two experiments together. Due to a combination of different limiting magnitudes and pixel resolutions, ATLAS and ASAS-SN light curves may behave differently even after the data cuts discussed in Section \ref{sec:data} have been imposed. We have visually inspected each light curve and chosen to show either ATLAS or ASAS-SN based on the quality of the data, the exceptions are:
\begin{itemize}
    \item omi Cet. We use American Association of Variable Star Observers (AAVSO) data\footnote{\url{https://www.aavso.org/data-download}} as it is too bright for ATLAS and ASAS-SN. We used the \textit{Vis.} band and cut upper limits.
    \item K 5-33 and WISE J192140.40+155354.6. No optical light curves as both ATLAS and ASAS-SN data are too noisy and sparse.
    \item IRAS 20124+1154. We include both to highlight an issue in ASAS-SN data that ATLAS can provide a more accurate alternative, see Section \ref{sec:suspected_d'} for more information.
\end{itemize}

\subsection{Confirmed D-types}\label{sec:confirmed_d}

\subsubsection{omi Cet = Omicron Ceti = Mira AB}
Omicron Ceti, more famously known as Mira AB, is a binary system containing the prototypical Mira variable star with recorded observations dating back to the 16th century. Owing to its close proximity of 110$\pm$18 pc \citep{gromadzki09} and wide binary separation of $\ge$70 AU, multi-wavelength studies have been able to resolve the two components of the system, finding evidence of Roche-lobe overflow (RLOF) in addition to wind accretion \citep{karovska06}\footnote{Rather than classical RLOF, where the donor's photosphere swells to fill the Roche lobe, Mira AB has a binary separation far too large for this to occur and instead can be interpreted as wind Roche-lobe overflow (WRLOF). In WRLOF, the dense AGB wind effectively fills the Roche radius, funneling mass transfer through the inner Lagrangian $L_1$ point to the accreting secondary \citep{abate13}.}. \citet{karovska05} found a soft X-ray outburst in late 2003 that may be associated with a stellar flare and subsequent mass ejection with a rough timescale of weeks to months, which could change the SED of both components for months to years.  This could, in turn, have led to a significant increase in dust formation ca. 2004/2005 and could have renewed the accretion disk around the WD. \citet{templeton09} present a $\sim$170 year visual light curve from 1838 through 2006, and though the study is focused on long-period variability, they make no mention of any noticeable difference in the 2004/2005 period.

\subsubsection{V852 Cen = Hen 2-104}
Hen 2-104, also known as the southern Crab nebula, is a Mira embedded within a bipolar nebula showing strong dust emission \citep{whitelock87}. The complex morphology of this nebula has been the target of many studies, with \citet{balick22} finding the nebular gas in both the ``inner hourglass'' and ``outer hourglass'' structures to be shock ionized by fast stellar winds rather than UV from the hot companion. \citet{santandergarcia08} find the estimated ionized mass of the nebula to be remarkably high at $\sim$0.16 M$_\odot$, indicating the main donor is the Mira. Bipolar jets along with fast winds from the WD could explain the extended structure, and they go on to note that the outflow speeds are slowest at low latitudes and highest in the polar direction which is the same as most bipolar PNe, suggesting the mechanism of formation is neither unique to SySts or PNe but could be common in high mass-loss rate systems.

\subsubsection{K 5-33}
K 5-33 was first classified as a PN (PBOZ 10) using IRAS and VLA measurements \citep{pottasch88}, but is shown to have Raman-scattered O\,{\sc vi} lines as presented in \citet{miszalski13}. They also go on to state that the OGLE-IV light curve has minimal variability which is consistent for a D-type SySt with an obscured Mira.

\subsubsection{H 1-36}
We mention in Section~\ref{sec:discussion} that H~1-36 may have excess mm/sub-mm emission that is explained by synchrotron radiation in the shock fronts of colliding winds. \citet{angeloni07} additionally add that broad IR lines suggest a high velocity component in the form of an X-ray jet may be present. The data \citet{angeloni07} used to perform this analysis span multiple decades, and they confirm that H~1-36 has not been significantly variable in the last 30 years. However, light curves presented in Figure~\ref{fig:d_lc_1} {\it do} show some variability over a $\sim$ 10-year period. It seems plausible, then, that the combination of the source variability at 95 GHz, and the large measurement uncertainties in the data of \citet{angeloni07} may be responsible for the apparent excess around 100 GHz reported for H~1-36.  If this is the case, then there is no need for a colliding wind mechanism to have a significant contribution from synchrotron radiation in H~1-36.

\citet{ivison94} note it as the first symbiotic OH/IR star.\footnote{OH stars are isolated Miras having copious FIR emission along with 1612 MHz OH masers.} They go on to explain that while molecular line emission and masers are uncommon in SySts due to photodissociation from the hot companion's UV radiation, that it is possible that dust shells can suppress this effect.
The ASAS-SN cameras in the \textit{g} filter have diverging values as seen in Figure~\ref{fig:d_lc_1}; we interpret this as due to diffraction spikes from the neighboring star G Sco ($\sim$78\arcsec separation, \textit{V} magnitude of $\sim$3.2) whose excessive brightness can clearly be seen in various photometric images in optical/IR bands.

\subsubsection{H 1-45}
\citet{miszalski13} find H 1-45 to have strong absorption bands of CN and Ba\,{\sc ii} $\lambda\lambda$ 4554, 4934, 6495, characteristic of a carbon star enhanced in s-process elements. Using the period-luminosity relation in the \textit{2MASS K$_S$} band for carbon Miras \citep{whitelock08}, they find a distance of 6.2$\pm$1.4 kpc, likely placing it on the near side of the Galactic Bulge. Though still possibly within the Galactic Disk, this would be the first detected carbon star within the Bulge and could shed insight on the longstanding problem of a lack of carbon stars in the Galactic Bulge \citep{feast07}.  

\subsubsection{H 2-38}
\citet{sanduleak73} classify H 2-38 as a ``Z-'' star, indicating that it has Z~Andromedae-like emission spectrum (sharp He\,{\sc ii} $\lambda$4686 line in addition to strong hydrogen emission plus weak He\,{\sc i}, and forbidden nebular lines) while ``-'' indicates that $\lambda$4686 is present but weaker than H$\beta$. They go on to note that ``Z-'' stars share characteristics with Wolf-Rayet stars and PNe, though H 2-38 is a confirmed SySt featuring Raman-scattered O\,{\sc vi} lines \citep{allen84b}.

\subsubsection{Hen 2-375}
Hen 2-375 shows O\,{\sc vi} features along with strong [O\,{\sc iii}] emission \citep{miszalski13}. No [N\,{\sc ii}] emission lines were detected in the nebula which is remarked as unusual, as symbiotic nebulae are typically high-density regions that show strong [N\,{\sc ii}] lines, and are a typical diagnostic used in separating SySts from PNe \citep{ilkiewicz17}. An extended blue nebula was observed, and the authors re-imaged Hen 2-375 with multiple filters to rule out photographic flaws. This blue nebula is supported by Hen 2-375's de-reddened G$_{\textrm{BP}}$-G$_{\textrm{RP}}$ color of $\sim$0.04 on the Gaia CMD in Figure~\ref{fig:cmd}.

\subsubsection{K 3-9}
\citet{ivison95a} describe K 3-9 to be a strong radio source, with data that match well to FF emission in the STB model. They find a sub-mm excess which they interpret as coming from cold dust ($\sim$30 K).  The source of dust is unclear: it could either be from mass loss by the WD or the Mira or produced during SyN outburst(s). \citet{munari02} refute the SyN possibility using optical (\textit{B, V}) light curves. A faint decrease in \textit{B} (17.2 mag in 1959 to 18.3 mag in $\sim$1970) is stated to be ``more reminiscent of classical symbiotic stars with moderate active phases than of symbiotic Miras in outburst.''

\subsubsection{HM Sge}
HM Sge has shown multiple outbursts and is one of the most recent SyN, with \citet{dokuchaeva76} reporting on a ``new emission object'' that went from 16 mag to 11 mag over the course of April to September 1975. Full light curves from AAVSO show a peak ca. 1981 with a steady decrease for $\sim$30 years until 2010. \citet{goldman24} present the \textit{B}, \textit{V}, \textit{I}, and \textit{R} AAVSO data showing a divergence in the light curves: \textit{B}, \textit{V}, and \textit{R} remain steady while \textit{I} shows an increase independent of the also-present periodic variability until 2022 where a small, brief outburst is experienced in all four filters followed by a steep decay. To explain this divergent behavior they give two suggestions: it could be related to the orbital motion of the system or it could be related to a dust obscuration event in the mid 1980's. \citet{munari89} observed a dimming in the \textit{J}, and \textit{K} bands starting in $\sim$1984 and continuing up to 1988, where their data end. They explain this as dust formation being inhibited by radiation from the WD, except in the Mira's ``shadow cone,'' which then passes into the line of sight of the observer during the binary orbit. 220~GHz luminosity shows an increase around the same time as the optical/NIR divergence described by \citet{goldman24} (Figure \ref{fig:lum_since_outburst}), possibly indicating dust formation. Additionally, 95, 150, and 220~GHz luminosities all show an increase around the same time as the 2022 outburst previously mentioned.

\subsubsection{RR Tel}
\citet{fleming08} were the first to discover the variability of RR Tel, noting a variation between 9 and 11.5 mag occurred between 1894 and 1907. \citet{dekock48} then found it to be a constant 7.4 mag in late 1946, and increased in brightness to 7.0 in March 1948 and 6.0 in July 1948, with \citet{mayall49} showing a light curve that captures the nova in outburst in late 1944 when the magnitude suddenly increased from 14 to 7. Although it has decayed to pre-nova brightness levels of ~$\sim$12 mag, there is still some variability present in ASAS-SN data. Short term spikes in optical data ($\sim$0.2 mag) are visible and relate to the Mira pulsation period of 386.73 days, while there is also either long period variability ($\sim$0.5 mag from 2017 to 2026) occurring or a continued decay from the nova. RR Tel was also observed in 2009 by SPT-SZ, the first generation camera on the SPT \citep{everett20}, and we include that 95, 150, and 220~GHz data in Figure~\ref{fig:lum_since_outburst}. 

\subsection{Confirmed D'-types}\label{sec:confirmed_d'}

\subsubsection{AS 201}
Two distinct morphological features are clearly seen in the H$\alpha$+[N\,{\sc ii}] and [O\,{\sc iii}] imaging of AS 201 as presented by \citet{schwarz91}. Both images clearly show the central star but a ring is seen in H$\alpha$+[N\,{\sc ii}]. This ring is interpreted as a low-density, low-excitation [N\,{\sc ii}] nebula that is the remnant of a PNe.  The high-density, high-excitation region is typical of SySts.  Using high resolution spectroscopy, \citet{pereira05} find the RG in AS 201 to show broadened absorption features due to rapid rotation, along with solar compositions of non-s-process elements. Ba and Y, which are s-process elements, have overabundances confirming AS 201 as a yellow symbiotic.

\subsubsection{ShWi 5}
ShWi 5 is noted to be a CN star with a temperature equivalent to a late G-type star.  Its 2MASS colors, a rich spectrum (including He\,{\sc ii}, [Fe\,{\sc vii}], [O\,{\sc iii}]), and a flat light curve are all typical for D'-type symbiotics \citep{miszalski13}. While TiO is not present in the giant continuum, it shows CN and CH features indicating it is carbon-rich.

\subsection{Confirmed S-types}\label{sec:confirmed_s}

\subsubsection{CN Cha}
CN Cha is an S-type SySt that had a nova outburst starting around 2012/2013. This nova is of particular interest due to the rarity of its evolution: a long-lasting flat peak at optical wavelengths of three years. \citet{kato23} provide light curve models comparing it to the symbiotic nova PU Vul --- which experienced a similar flat peak optical nova of eight years in 1979 --- as well as the more common fast rise/sharp decay novae such as those seen in V1500 Cyg, V838 Her, and V1668 Cyg. 

A very thorough search of archival data including photometry, light curves, and spectroscopy from IR to UV is presented by \citet{lancaster2020}. They show SEDs before and after the event, overlaying the same blackbody fits and provide an example of how spectra can dramatically change over time and that near simultaneous observations are necessary when trying to fit any potential emission models.

\subsubsection{V455 Sco}
\citet{fekel08} provide a detailed history of V455 Sco in addition to analysis of IR spectroscopy that yields orbital parameters. They find the red giant in V455 Sco to be an AGB based on its luminosity, and the system to be high-mass, with M$_{WD}$ = 1.05 M$_{\odot}$ and M$_{AGB}$ = 2.9 M$_{\odot}$. They determine that the system is eclipsing, with an inclination, \textit{i}, of 94{\textdegree}$\pm$1{\textdegree}, a semi-major axis, \textit{a}, of 627.6 R$_{\odot}$, and an orbital period of 1439 days. The period is in agreement with inferences based on a fragmented light curve spanning over 50 years from approximately 1885-1938 \citep{swope40}. While the eclipse was observed in the 1920s, there is an unexpected brightening around 1933 of \textgreater2 mag compared to its typical maximum which lasts multiple years, starting to decay right as the data ends. We confirm the periodic variability due to the orbit in ASAS-SN and ATLAS data in Figure~\ref{fig:s_lc}, but we see no evidence of any recent outbursts so we tentatively include it as an outbursting system in Table~\ref{tbl:systs_main}. 

\citet{akras19} list V455 Sco as a fourth classification of SySts, an ``S+IR'' type: S-type SySts with an unexplained FIR excess. \citet{merc22} investigate the individual S+IR type SySts and are able to explain 35 of 37 as some combination of variable/unreliable IR data, objects misclassified as SySts, and incorrect temperatures inferred that result in reclassification as D'-types. Of their remaining two SySts that appear to have actual excess IR in WISE and AKARI data, one is V455 Sco, though they neglect to include these as a special category due to the otherwise heterogeneous nature of the SySt population. In Figure~\ref{fig:s_lc} we note the sparse availability of \textit{W1} and \textit{W2} data, though it has increased from 6 mag to 5 mag (from $\sim1\times10^3$ to $3\times10^3$ mJy) in AllWISE and NEOWISE data between 2010 and 2023. As mentioned previously, V455 Sco has evidence of a historical outburst of unknown origin; it is unclear if a similar outburst may have recently occurred that could explain this IR excess. We note that NODSV-provided AKARI and WISE data are in agreement, while IRAS 12 and 25 $\mu$m data are roughly 2-3 times higher which could indicate significant variability at those wavelengths.

\subsubsection{AR Pav}
AR Pav has an erratic light curve dating back to 1889, owing to its eclipsing binary nature and numerous Z~And outbursts \citep{mayall37}. An analysis of the system and modeling of a century of the historical light curve is provided by \citet{skopal00}. They note the quiescent intervals have similar features to those of cataclysmic variables (CVs), but their model of an accretion disk around the hot component does not explain the energy balance of emission and suggests a different geometry of mass transfer than those of CVs. The most recent recorded outburst occurred in 2021 \citep{merc21}, and while the light curve in Figure~\ref{fig:s_lc} shows more recent activity, this appears to be orbital phase-locked activity related to eclipsing and the $\sim$605 day orbital period \citep{fekel17}.

\subsubsection{DD Mic}
DD Mic, also referred to as CD-43{\textdegree}14304, is an S-type yellow symbiotic \citep{pereira09}. It has a complex light curve with a tenuous history of outbursts; \citet{gromadzki09} note it as having eclipse-like minima, variations with orbital period, and showing outbursts. They describe periodic brightenings as an enhanced accretion rate during periastron passage that occur a few hundred days after the periastron, consistent with an orbital period of $\sim$1450 days \citep{schmid98}. \citet{skopal05} present a brief history of DD Mic, and their analysis supports the variations in optical and far-UV being due to orbital modulations. Mild brightenings in ASAS-SN and WISE data around 2019 and 2023 can be seen in Figure~\ref{fig:s_lc}, which correspond to the roughly four year orbital period. We do not include DD Mic as having an outburst in Table~\ref{tbl:systs_main}, as periastron passage variability is not in the same family of triggered thermonuclear ignition as Z~And/SyN/SyRN (or type-Ia SNe) outbursts.

\subsubsection{AG Peg}
AG Peg is the oldest known SyN, having recorded observations going back to the 1800s. \citet{fleming94, cannon16} noted the bright hydrogen lines and relation to P Cygni type stars, while \citet{merrill16} remarked that future investigations could possibly relate it to novae. Inspired by these observations, \citet{lundmark12} did an archival search of catalogs dating back to 1821 and found an increase in magnitude from 9 to 6.3 occurred between 1841 and 1871. \citet{tomov16}, and the sources within, continue describing the observational history of AG Peg, including misclassified ``abrupt changes of the star brightness'' in the 20th century that can be attributed to the orbital period; however, their main point is to report on an outburst in late 2015 that is distinctly different than any orbital modulation and classify it as a Z~And outburst due to the behavior of the O\,{\sc vi} $\lambda$6825 emission.

\subsubsection{R Aqr}
R Aqr is the nearest and best-studied SySt. It hosts a Mira variable giant with a pulsation period of 388 days, along with an extended nebula that has been imaged by HST \citep{burgarella92} and ALMA \citep{gomez24}. The  jets and nebula in R Aqr are thought to result from outbursts, of which there are multiple credible claims throughout history. Korean astronomers detected a star brightening near the Aquarii constellation in 1073 and 1074 \citep{yang05}, with nitrate ion concentrations in antarctic ice core samples potentially giving corroborating evidence \citep{tanabe12}. Initial measurements of expansion of the nebula by Hubble and Baade indicated that it was ejected 600 years prior \citep{adams44} while \citet{solf85} provide evidence of both the outer and inner nebular shells being the results of outbursts separated by $\sim$450 years (ca. 1340 and 1800). Eclipses of R Aqr have been observed in the 1930s and 1970s, with \citet{wilson81} finding an orbital period of 44 years and eclipse duration of 8.5 years. They predicted the next eclipse would occur between 2018-2026 which is clearly visible in the ASAS-SN light curve in Figure~\ref{fig:s_lc}.

\subsection{Suspected D-types}\label{sec:suspected_d}

\subsubsection{ASASSN-17dm}
ASASSN-17dm was originally thought to be a supernova when first detected in February 2017, increasing in brightness from \textit{V} magnitude \>17.1 to 15.9 over a timespan of about a month \citep{kiyota17}. Follow-up spectroscopy detected an M-type star in the system \citep{morrell17} while \citet{fraser17} also notes the ``flux is most likely from a nebula around a Galactic source (e.g. nova, or symbiotic star) and the system is in outburst.'' This outburst is classified as a Z~And outburst in light curve analysis using a random forest classifier in the ASAS-SN Catalog of Variable Stars \citep{jayasinghe19}.

\subsubsection{IRAS 18344-0632 = G25.5+0.2}
There is a lot of uncertainty behind the nature of G25.5+0.2, having previously been categorized as an H\,{\sc ii} region, PN, young stellar object, supernova remnant, and luminous blue variable \citep{phillips08}. Using MIR images from Spitzer, \citet{phillips08} identify a change in morphology from 3.6--8~$\mu$m, with shorter wavelengths being dominated by the central star and showing a bilobal structure while longer wavelengths reveal a double-peaked nebula. Using archival data, they present an IR through radio SED, estimating $\nu_t \sim$1 GHz which is in agreement with previous studies and a peak around 25-70$\mu$m though stating a lack of photometry between 70$\mu$m and 1 cm prevents them from  properly defining this peak. They also find variability of the central star in 2MASS data, to the extent that would be unusual for a PN but does not necessarily rule that possibility out. After going through the possibilities of previous classifications they suggest a collimated outflow encircled by a dusty torus and lean towards the cause being symbiotic outflow of a D-type due to the previously mentioned central star variability. 
We note that purely based on IR classifications the SED peak at $\sim$25--70 $\mu$m should make this a D'-type, though variability as well as very high reddening at shorter wavelengths -- we find $E(B-V)$ $\sim$ 6.61 using \texttt{mwdust} -- can lead to inaccuracies in that approximation. \citet{akras19} maintain the D-type possibility, assigning it an SED peak at 3.6 $\mu$m.

\subsubsection{PM 1-253}
\begin{figure*}[h]
    \centering
    \includegraphics[width=\textwidth,center]{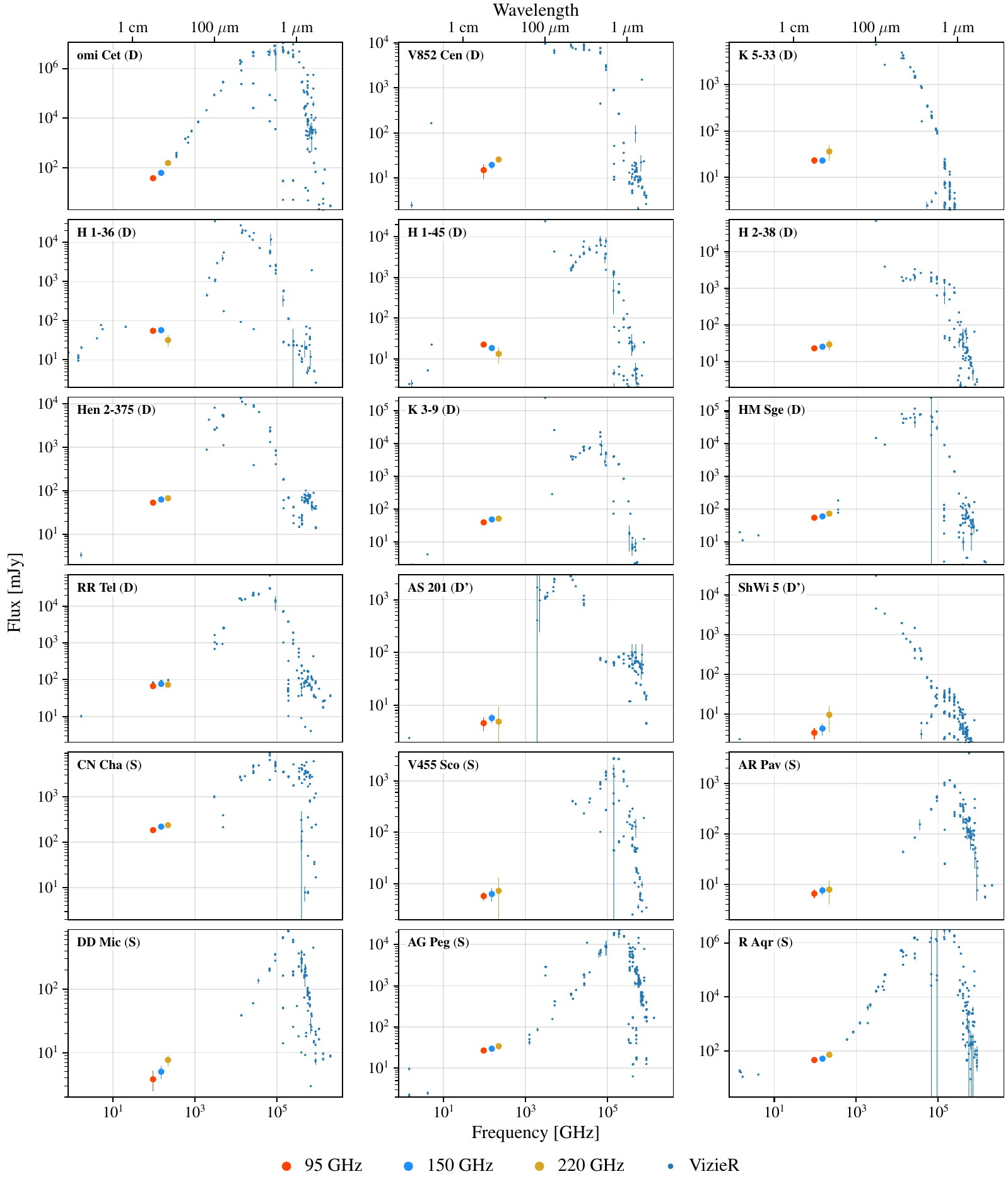}
    \caption{SEDs of all confirmed SySts. SySts detected by both SPT and ACT use flux values from the higher SNR detection.}
    \label{fig:confirmed_sed}
\end{figure*}

\begin{figure*}[h]
    \centering
    \includegraphics[width=\textwidth,center]{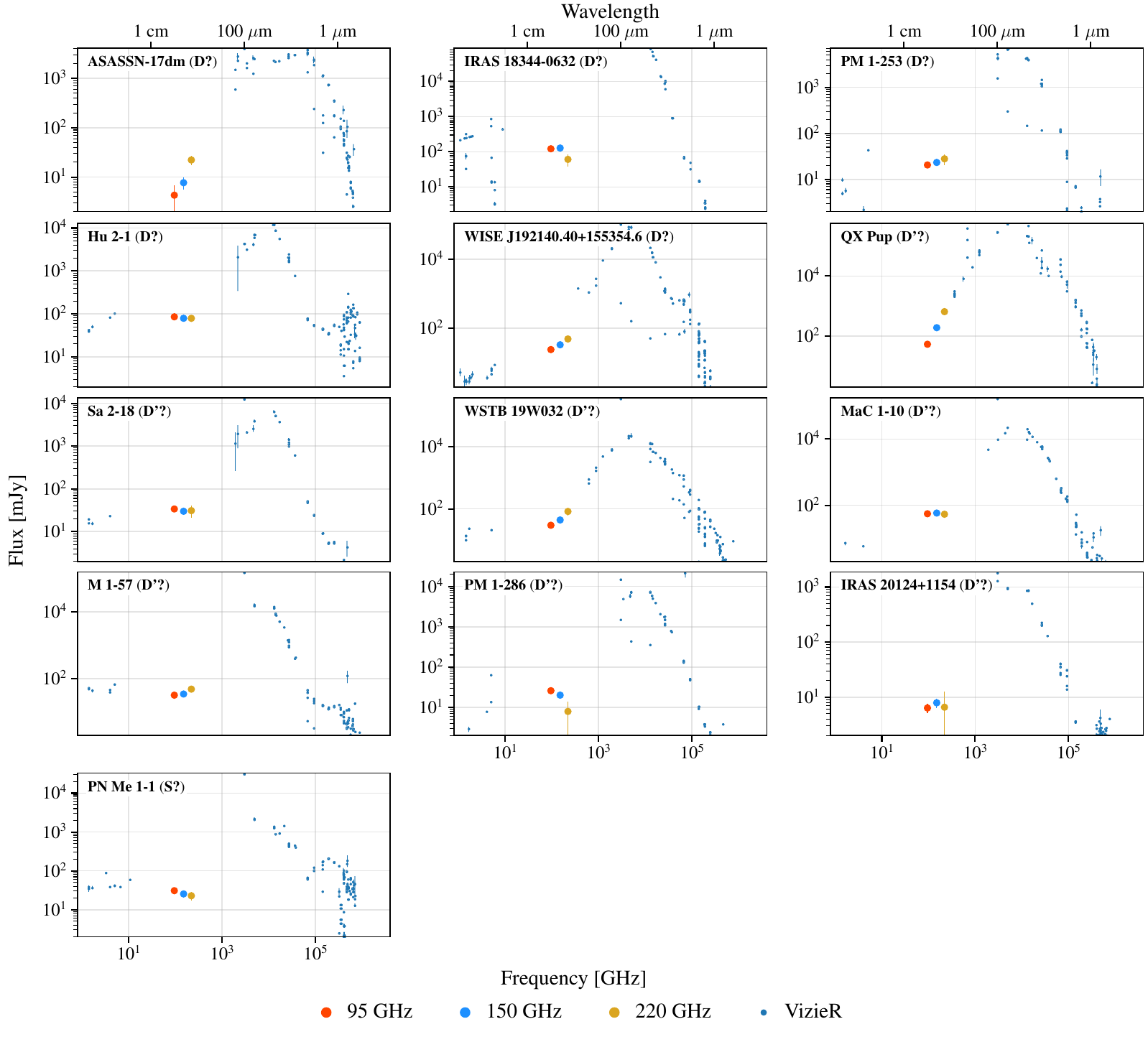}
    \caption{SEDs of all suspected SySts. SySts detected by both SPT and ACT use flux values from the higher SNR detection.}
    \label{fig:suspected_sed}
\end{figure*}

\begin{figure*}
    \includegraphics[width=.99\textwidth]{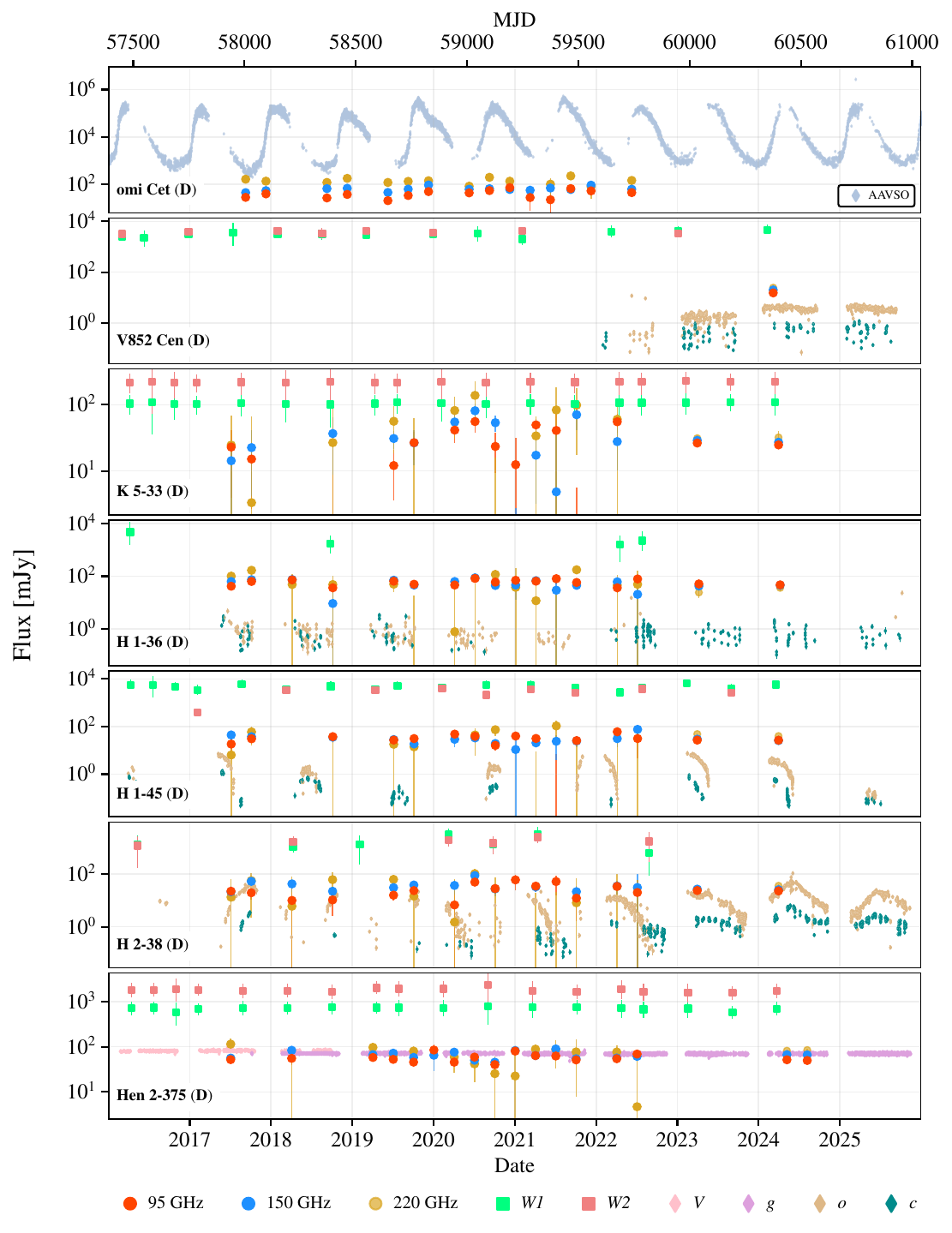}
    \caption{Multi-wavelength light curves of D- and D'-type SySts (Figure 1 of 2).}
    \label{fig:d_lc_1}
\end{figure*}

\begin{figure*}
    \includegraphics[width=.99\textwidth]{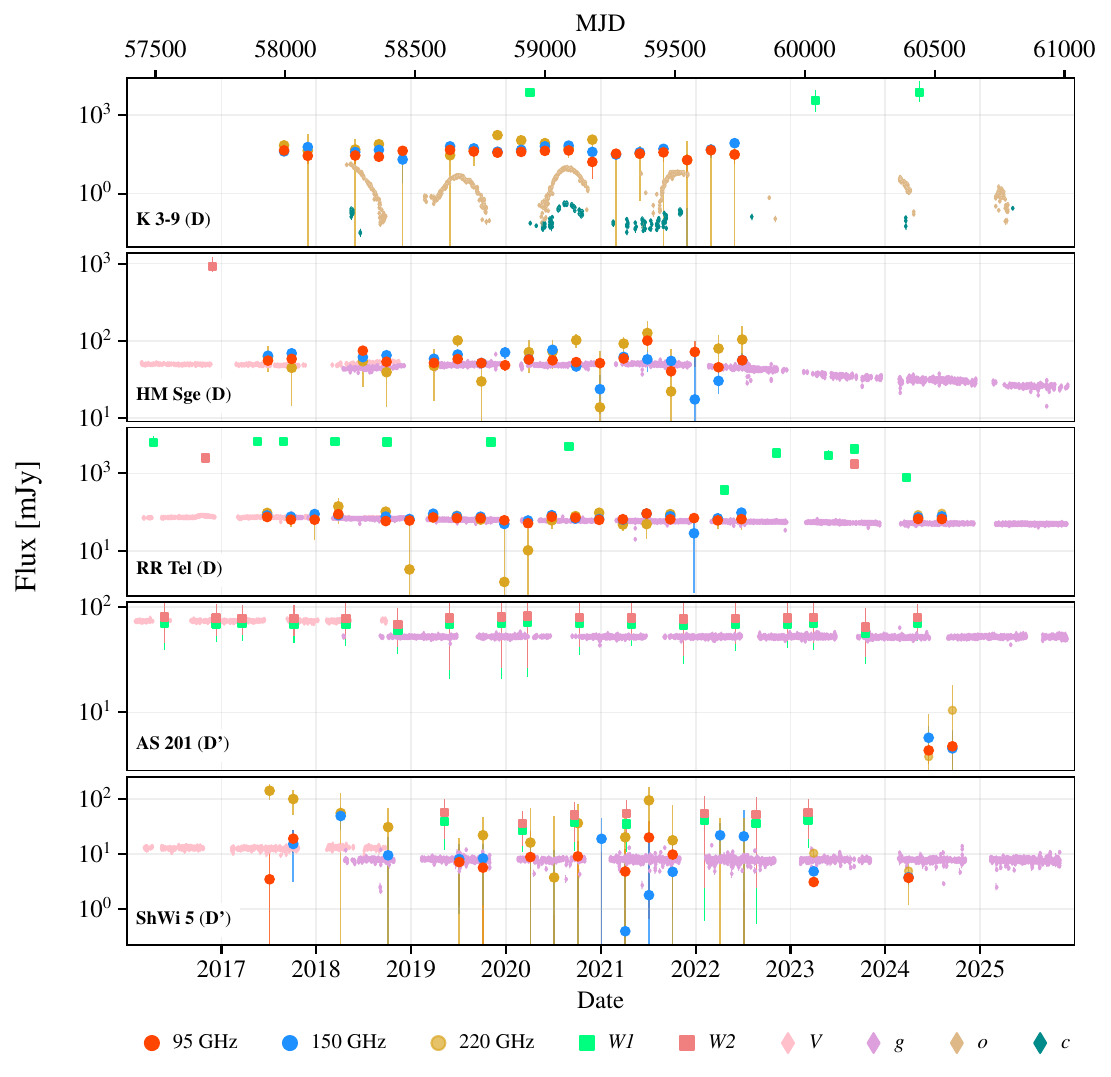}
    \caption{Multi-wavelength light curves of D- and D'-type SySts (Figure 2 of 2).}
    \label{fig:d_lc_2}
\end{figure*}

\begin{figure*}
    \includegraphics[width=.99\textwidth]{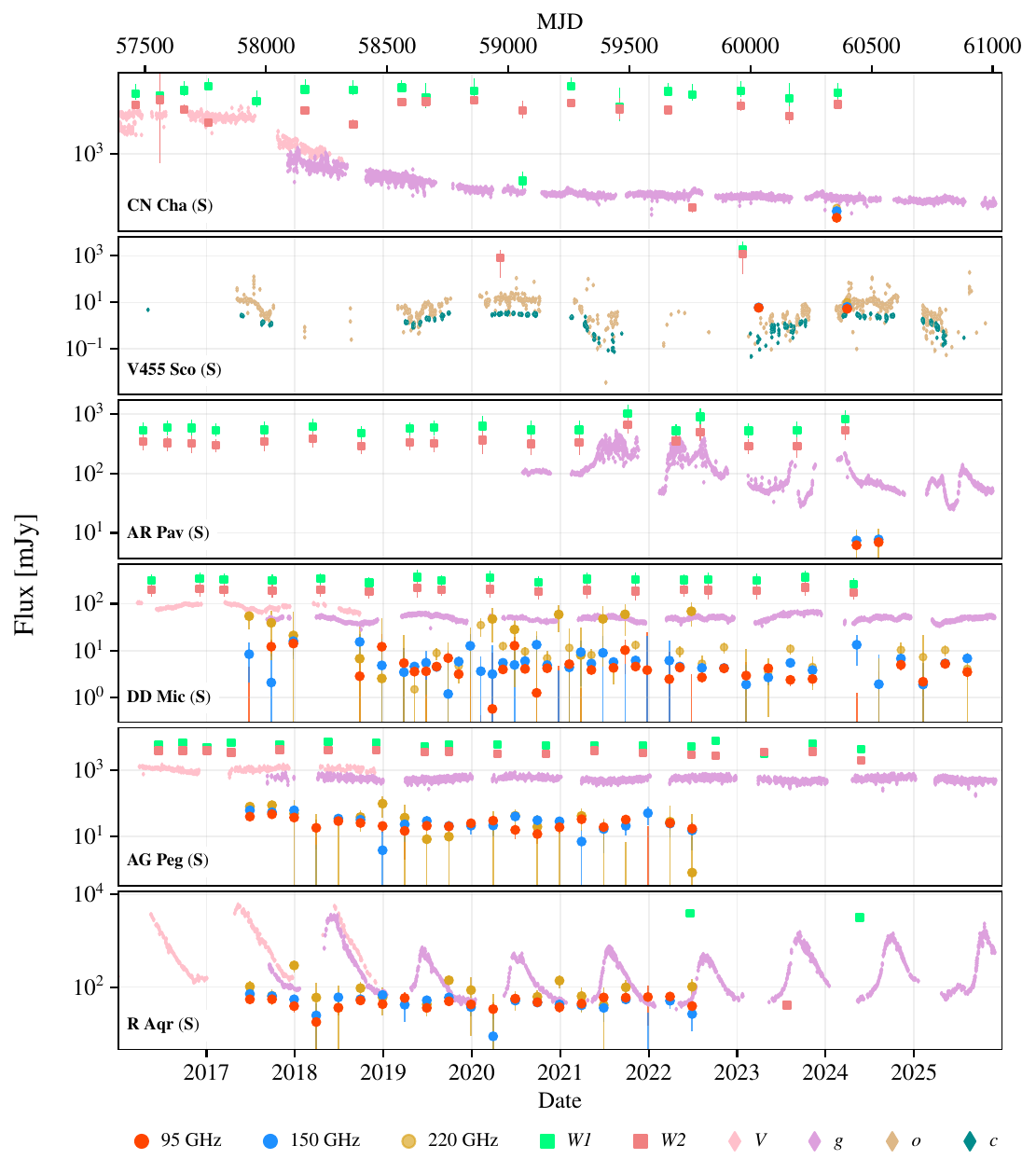}
    \caption{Multi-wavelength light curves of S-type SySts.}
    \label{fig:s_lc}
\end{figure*}

\begin{figure*}
    \includegraphics[width=.99\textwidth]{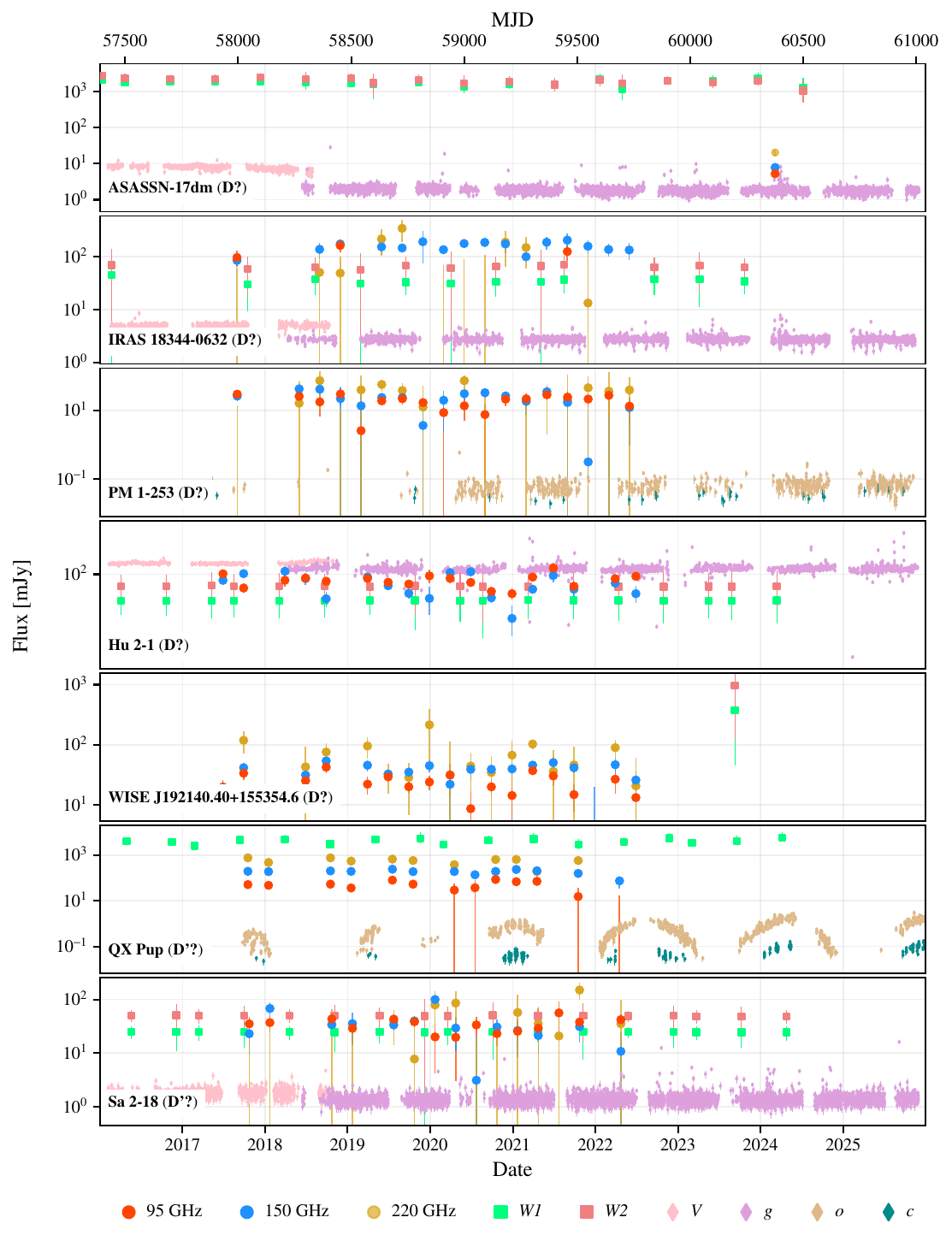}
    \caption{Multi-wavelength light curves of suspected SySts (Figure 1 of 2).}
    \label{fig:susp_lc_1}
\end{figure*}

PM 1-253 is noted as having highly collimated bipolar outflows in HASH \citep{parker16}. Spectroscopy from \citet{vandesteene96} shows strong H$\alpha$ and [N\,{\sc ii}] $\lambda\lambda$ 6548,6583 lines as well as H$\beta$ and H$\gamma$, with a marginal detection of [O\,{\sc iii}] $\lambda$4959 and no detections of He\,{\sc ii} $\lambda$4686 and [O\,{\sc iii}]. \citet{miranda07} find [O\,{\sc iii}] missing in the nebula but present around the central star as well as other differences between the lobes and central region. Using classification tree models on various IR color criteria to determine probability of genuine PNe, \citet{akras19b} state that PM 1-253 (referred to as IPHASXJ184336.6+034640) ``is classified as a probable SySt but it satisfies the criteria of PN and not those of SySts. It has a very high W1-W4 colour index, which is indicative of a genuine PN.'' We note that PM 1-253 (Figure~\ref{fig:suspected_sed}) has a very similar NIR/MIR spectra to K 5-33 (Figure~\ref{fig:confirmed_sed}) -- a confirmed D-type SySt that shows Raman-scattered O\,{\sc vi} lines -- though we do not assume this to mean PM 1-253 is conclusively a SySt.

\subsubsection{Hu 2-1}
\citet{miranda01} identify various morphological features in Hu 2-1 using [N\,{\sc ii}] $\lambda$6584 and H$\alpha$ images from the \textit{Hubble Space Telescope} (HST), including a bipolar nebula with a point-symmetric inner shell nested inside as well as smaller features indicating collimated outflows and possible bow-shock structures. The morphology, along with first-order estimates of orbital parameters, show many similarities to confirmed SySts R Aqr and HM Sge. They suggest Hu 2-1 is a SySt where the AGB central star has evolved into a PNe.

\subsubsection{WISE J192140.40+155354.6}
Very little is known about WISE J192140.40+155354.6. Using the INT Photometric H-alpha Survey (IPHAS) catalog \citep{gonzalez08}, a semi-automated search for PNe was conducted by \citet{viironen09}. They required detections in the H$\alpha$ and Sloan \textit{r'} filters and then classified candidate PNe based on IPHAS and 2MASS color-color diagrams. They describe WISE J192140.40+155354.6 as a butterfly shaped nebula and preliminarily classify it as a SySt based on spectra from the Macquarie/AAO/Strasbourg H$\alpha$ Planetary Nebula Catalog (MASH). We note the spectra of this object do not provide a strong argument for the SySt classification, though a reddening of $E(B-V)=5.99$ \citep{schlafly11} makes any definitive classification challenging. 

\subsection{Suspected D'-types}\label{sec:suspected_d'}

\subsubsection{QX Pup}
The ``rotten egg'' or ``Calabash'' nebula, QX Pup (also referred to as OH 231.8+4.2) has a contentious history despite being well studied: being classified as a SySt, proto-planetary nebula (PPN) or an M9-10 III star with an A-type main sequence (MS) companion often between the same authors \citep{alcolea96, sanchez04}. \citet{alcolea01} give insight into the confusion, stating that the massive molecular nebula is unlike any other seen in SySts, while the structure of the lobes is unlike any seen in PPNe. Using the Atacama Large Millimeter/submillimeter Array (ALMA), \citet{sanchez22} explore the structure of QX Pup in detail, noting that the process that formed the nebula $\sim$800 yr ago required accretion from a compact object and subsequent jets, referring to the MS companion in \citet{sanchez04}, and that the current dynamics of the system are no longer in an ''active'' state and could be in a quiescent state similar to those of SySts. \citet{frew10}, in their summary of PNe discovery techniques and mimics, claim QX Pup is probably a SySt due to the 0.4 pc size of its nebula along with the presence of a heavily obscured Mira, and should be removed from the PPN class.

\subsubsection{Sa 2-18}
Very little is known about Sa 2-18. Classified as a likely SySt in HASH \citep{parker16}, it shows typical SySt signatures of [O\,{\sc iii}], H$\beta$, He\,{\sc ii}, and [O\,{\sc iii}]/H$\gamma$. It is included in \citet{luo05} as a detected PNe with a 1.4 GHz flux of 17.2 mJy.

\subsubsection{WSTB 19W032 = 19W32}
19W32 is described as a narrow-waist bipolar nebula with a rising spectral index at radio wavelengths of 0.7 and 1.3~cm \citep{lee07}. They find an elongated radio core that is larger than the predicted source size and a high derived mass loss rate from the wind, which they state can be attributed to a collimated wind or jet possibly due to a binary system. \citet{schmeja01} produce a de-reddened NIR \textit{I}, \textit{J}, and \textit{K} color--color diagram and find that 19W32 is separate from both the PNe and symbiotic Mira populations and is placed alongside Miras and red semi-regular variables, noting that while the extinction is uncertain it is not a genuine PN and could possibly be a SySt.

\subsubsection{MaC 1-10}
\citet{gorny04} classifies this object as a Wolf-Rayet PN based on relative intensities of C\,{\sc iii} $\lambda$5695 and C\,{\sc iv} $\lambda$5805 taken from a sample of IRAS color-selected stars near the Galactic center. HASH classifies it as a SySt candidate citing the [O\,{\sc iii}]/H$\gamma$ and [N\,{\sc ii}]/H$\alpha$ ratios, and notes there are highly collimated bipolar flows.  

MaC 1-10 shows water maser emission \citep{2008AJ....135.2074G}, as well.  It is plausible that these are connected, and that the mm-wave emission in this system, as in many other Wolf-Rayet PNe, is primarily from synchrotron emission due to shocks.  There are some indications that this can be a transient phenomenon \citep{2024A&A...688L..21H}, so continued variability analysis on this object in the SPT Galactic Plane Survey project will be interesting.  

\subsubsection{M 1-57}
Not much is known about M 1-57. It is listed in HASH as a high excitation bipolar PNe \citep{parker16}, indicating a close binary where a companion has disrupted the outflow of material from the central star. Spectroscopy is provided from \citet{kwitter01} showing emission lines of [O\,{\sc iii}] He\,{\sc ii}, H$\beta$, and H$\gamma$ that are in general agreement with what is expected to be seen in a SySt. [N\,{\sc ii}]/H$\alpha$ and [S\,{\sc ii}] are noted to indicate high density.

\subsubsection{PM 1-286}
Not much is known about PM 1-286. Using H$\alpha$/H$\beta$ ratios, \citet{vandesteene96} find high visual extinction of $A_V=8.7$ and claim to see some [Fe\,{\sc i}], [Fe\,{\sc ii}], [Fe\,{\sc iv}], and [Fe\,{\sc vii}] emission lines. A table of line flux ratios is provided, and comparing [O\,{\sc iii}] to hydrogen ratios ($\lambda$4363/H$\gamma$, $\lambda$5007/H$\beta$) place it firmly in the SySt population shown in Figure~3 of \citet{pereira05b}.

The source shows evidence for a strongly peaked spectral energy distribution in the radio-through-mm bands.  It is detected at about 3.3 mJy in RACS-Mid and RACS-High, at 1.4 and 1.7 GHz, but not seen in RACS-Low at 888 MHz \citep{2025PASA...42...38D}.  It is also seen in VLASS at 3 GHz \citep{2021ApJS..255...30G}, at 7.8$\pm$0.3 mJy, and in the RMS Survey at 13.9$\pm$0.3 mJy at 5 GHz \citep{2009A&A...501..539U}, but then shows a spectral index of $-0.6\pm0.2$ between 95 and 150~GHz in the ACT data.  The source is consistent with being pointlike in the radio bands \citep{2009A&A...501..539U}.  These properties are probably easier to explain in terms of colliding winds in a young planetary nebula with a binary companion, so that the whole spectrum is synchrotron emission, and is self-absorbed below some frequency between 5 and 95 GHz. 

\subsubsection{IRAS 20124+1154 = PM 1-322}
PM 1-322 was spectroscopically identified as a PNe in 2005 by \citet{pereira05b}, even though a diagnostic plot comparing [O\,{\sc iii}] to hydrogen ratios ($\lambda$4363/H$\gamma$, $\lambda$5007/H$\beta$) firmly places it in the SySt population, in the region of the D and D'-types. The lack of absorption lines and TiO bands along with the lack of a red continuum excess (indicating a companion) was explained as PM 1-322 being a young high-density PNe. \citet{akras19} classifies it as a D'-type in a catalog classifying SySts by their SEDs using AllWISE and 2MASS data and confirming with Gaia derived temperatures.

\begin{figure*}
    \includegraphics[width=.99\textwidth]{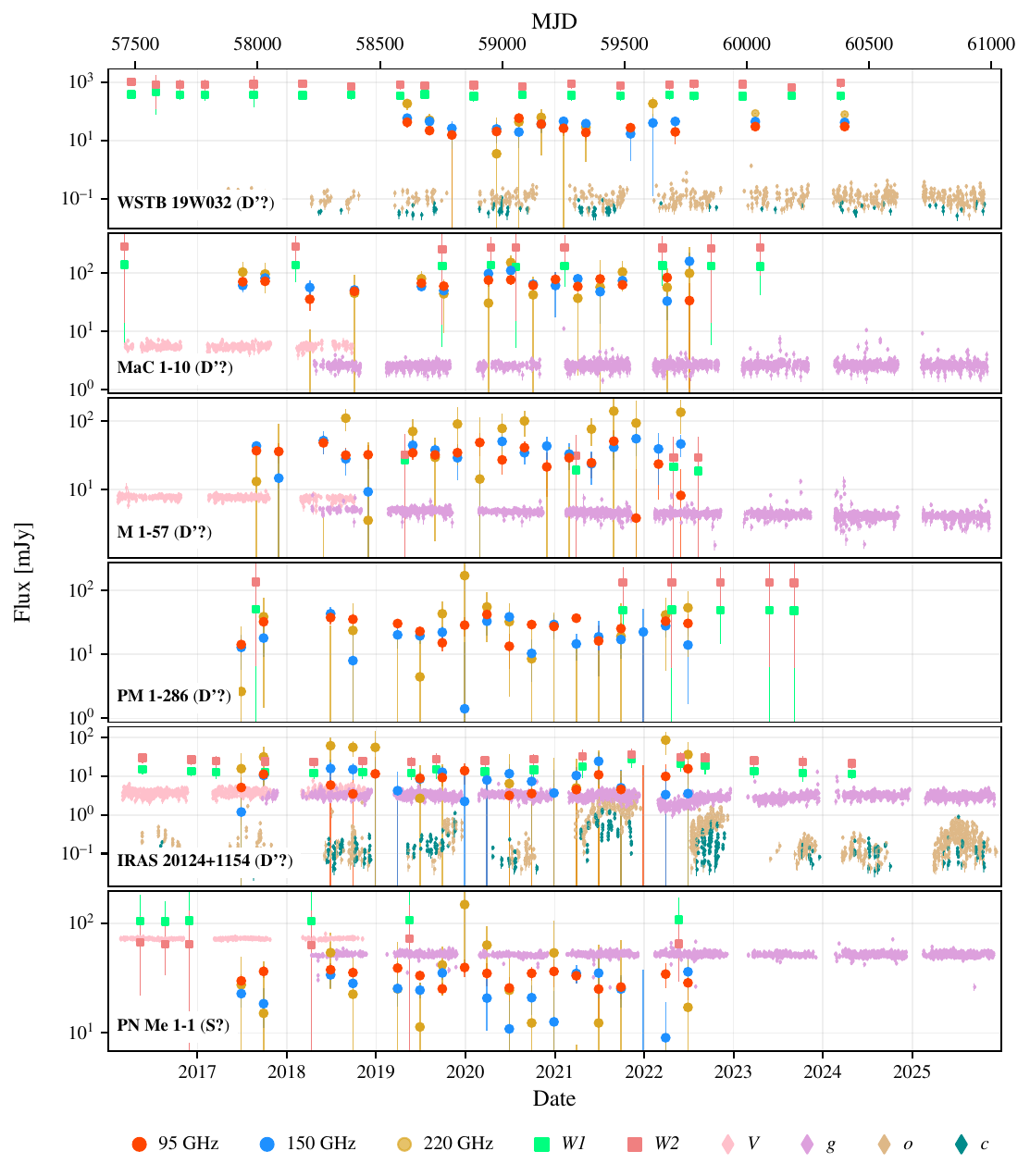}
    \caption{Multi-wavelength light curves of suspected SySts (Figure 2 of 2).}
    \label{fig:susp_lc_2}
\end{figure*}

\citet{paunzen23} investigates PM 1-322 thoroughly: constructing the SED, following up with low-resolution spectroscopy, and providing a multi-wavelength light curve with WISE, ASAS-SN, and ZTF data. They find multiple brightening events in the WISE \textit{W1} and \textit{W2} bands (roughly MJD~57200, 59270--59530) that are not as apparent in optical wavelengths. The MJD~57200 event only has coverage from ASAS-SN and is not seen in either the \textit{V} nor \textit{g} filters. The MJD 59270 event likewise is not seen by ASAS-SN, but has ZTF coverage and is weakly seen in the \textit{g} filter and strongly seen in the \textit{r} filter. However, these optical brightenings (along with another, smaller bump seen in WISE around MJD~58750) are clearly seen by ATLAS in both the \textit{o} and \textit{c} filters, shown in Figure~\ref{fig:susp_lc_2}. After the MJD~59270 brightening there is a drop in brightness of about 1 magnitude that lasts for roughly half a year, which they refer to as ``an eclipse.'' This is clearly seen in both ASAS-SN and ZTF, and while ATLAS does not have coverage of the start of this ``eclipse,'' there is plausible evidence of brightening from a lower magnitude to above a baseline value.

\citet{paunzen23} provide multiple interpretations for this seemingly odd behavior with the favored scenario being a ``puffed-up dusty disk:'' a central hot star with an inner hot gas disk and outer cool dust disk that are both seen edge-on. Both disks expand and get hotter for some reason, which would obscure the central star and show stronger dimming at shorter wavelengths due to dust extinction. A possibility for this is given as accretion from a cool companion, as seen in SySts in active shell-burning phases. This would obscure direct viewing of the hot companion but would allow ionizing light to reach the CSM producing strong emission lines. Given the new information that these events are actually seen at optical wavelengths, a re-interpretation of this system is encouraged.

\subsection{Suspected S-types}\label{sec:suspected_s}

\subsubsection{PN Me 1-1}
Originally classified as a PN, PN~Me~1-1 has been shown to contain a cool central star of type K1-4 II \citep{pereira08, shen04} embedded in a nebula. The temperature of such a star is not high enough to produce the ionization observed in the nebula, suggesting a faint hot binary companion which would classify it as a peculiar PNe  or possible a yellow symbiotic. 

\citet{shen04} go on to classify PN Me 1-1 as a yellow symbiotic based on strong emission lines (e.g. H\,{\sc i}, He\,{\sc i}, He\,{\sc ii}, [O\,{\sc iii}], [Ne\,{\sc iii}], etc.) along with ``a marginal unresolved feature'' at $\lambda$6825 which they go on to tentatively identify as the Raman-scattering of the [O\,{\sc vi}] $\lambda$1032 resonance line. Investigation of s-process elements, typical in yellow symbiotics, yielded an absorption feature at 6945~\AA, which is interpreted as a possible blending of Ba\,{\sc ii}, Ca\,{\sc i} and Fe\,{\sc i} but failed to find any other significant absorption lines of Ba\,{\sc ii}. 

\citet{pereira08} investigate the rotational velocity of the K bright giant in PN Me 1-1, with a mechanism for rapid rotation as spin-up due to the accretion of mass ejected from the hot companion when it was formerly an AGB. Using high resolution spectroscopy, they determine PN Me 1-1 is a rapid rotator with a lower limit on V$_{\mathrm{rot}}$ of $90\pm10$ km s$^{-1}$. Comparing Ba/Fe abundance ratios to D'-type SySts show that this abundance declines along with temperature, and go on to suggest the cool central star is actually an RGB with a massive convective envelope sufficient enough to dilute any accreted s-process elements as it evolved from the subgiant branch. While there is no conclusive evidence suggesting it as a bona fide symbiotic star, there is plausible evidence for the case.

\clearpage

\clearpage

\clearpage

\clearpage

\end{document}

%% file: authors.tex
\def\ILAst{1}
\def\PhysicsPrinceton{2}
\def\TTU{3}
\def\FNAL{4}
\def\KICPChicago{5}
\def\AAUChicago{6}
\def\Melbourne{7}
\def\IAP{8}
\def\UNM{9}
\def\ANLHEP{10}
\def\KIPAC{11}
\def\Stanford{12}
\def\SLAC{13}
\def\LMU{14}
\def\EFIChicago{15}
\def\PhysicsUChicago{16}
\def\Geneva{17}
\def\NTU{18}
\def\Berkeley{19}
\def\Saclay{20}
\def\KEK{21}
\def\McGill{22}
\def\CIFAR{23}
\def\USTCAst{24}
\def\USTCPhys{25}
\def\ILPhys{26}
\def\UCLA{27}
\def\MSU{28}
\def\Caltech{29}
\def\UCDavis{30}
\def\CASA{31}
\def\ColoradoPhys{32}
\def\ILNCSA{33}
\def\SKAI{34}
\def\CaseWestern{35}
\def\CfA{36}

\author{
  C.~Tandoi\altaffilmark{\ILAst}\orcidlink{0000-0002-2077-6004},
  A.~Foster\altaffilmark{\PhysicsPrinceton}\orcidlink{0000-0002-7145-1824},
  T.~J.~Maccarone\altaffilmark{\TTU}\orcidlink{0000-0003-0976-4755},
  A.~J.~Anderson\altaffilmark{\FNAL,\KICPChicago,\AAUChicago}\orcidlink{0000-0002-4435-4623},
  B.~Ansarinejad\altaffilmark{\Melbourne},
  M.~Archipley\altaffilmark{\AAUChicago,\KICPChicago}\orcidlink{0000-0002-0517-9842},
  L.~Balkenhol\altaffilmark{\IAP}\orcidlink{0000-0001-6899-1873},
  D.~R.~Barron\altaffilmark{\UNM}\orcidlink{0000-0002-1623-5651},
  K.~Benabed\altaffilmark{\IAP},
  A.~N.~Bender\altaffilmark{\ANLHEP,\KICPChicago,\AAUChicago}\orcidlink{0000-0001-5868-0748},
  B.~A.~Benson\altaffilmark{\FNAL,\KICPChicago,\AAUChicago}\orcidlink{0000-0002-5108-6823},
  F.~Bianchini\altaffilmark{\KIPAC,\Stanford,\SLAC}\orcidlink{0000-0003-4847-3483},
  L.~E.~Bleem\altaffilmark{\ANLHEP,\KICPChicago,\AAUChicago}\orcidlink{0000-0001-7665-5079},
  S.~Bocquet\altaffilmark{\LMU}\orcidlink{0000-0002-4900-805X},
  F.~R.~Bouchet\altaffilmark{\IAP}\orcidlink{0000-0002-8051-2924},
  E.~Camphuis\altaffilmark{\IAP}\orcidlink{0000-0003-3483-8461},
  M.~G.~Campitiello\altaffilmark{\ANLHEP},
  J.~E.~Carlstrom\altaffilmark{\KICPChicago,\EFIChicago,\PhysicsUChicago,\ANLHEP,\AAUChicago}\orcidlink{0000-0002-2044-7665},
  J.~Carron\altaffilmark{\Geneva}\orcidlink{0000-0002-5751-1392},
  C.~L.~Chang\altaffilmark{\ANLHEP,\KICPChicago,\AAUChicago},
  P.~M.~Chichura\altaffilmark{\PhysicsUChicago,\KICPChicago}\orcidlink{0000-0002-5397-9035},
  A.~Chokshi\altaffilmark{\AAUChicago},
  T.-L.~Chou\altaffilmark{\AAUChicago,\KICPChicago,\NTU}\orcidlink{0000-0002-3091-8790},
  A.~Coerver\altaffilmark{\Berkeley}\orcidlink{0000-0002-2707-1672},
  T.~M.~Crawford\altaffilmark{\AAUChicago,\KICPChicago}\orcidlink{0000-0001-9000-5013},
  C.~Daley\altaffilmark{\Saclay,\ILAst}\orcidlink{0000-0002-3760-2086},
  T.~de~Haan\altaffilmark{\KEK}\orcidlink{0000-0001-5105-9473},
  K.~R.~Dibert\altaffilmark{\AAUChicago,\KICPChicago},
  M.~A.~Dobbs\altaffilmark{\McGill,\CIFAR},
  M.~Doohan\altaffilmark{\Melbourne},
  D.~Dutcher\altaffilmark{\PhysicsPrinceton}\orcidlink{0000-0002-9962-2058},
  C.~Feng\altaffilmark{\USTCAst,\USTCPhys,\ILPhys},
  K.~R.~Ferguson\altaffilmark{\UCLA,\MSU}\orcidlink{0000-0002-4928-8813},
  N.~C.~Ferree\altaffilmark{\Caltech,\KIPAC,\Stanford}\orcidlink{0000-0002-7130-7099},
  K.~Fichman\altaffilmark{\PhysicsUChicago,\KICPChicago},
  S.~Galli\altaffilmark{\IAP},
  A.~E.~Gambrel\altaffilmark{\KICPChicago},
  A.~K.~Gao\altaffilmark{\ILPhys},
  F.~Ge\altaffilmark{\Caltech,\KIPAC,\Stanford,\UCDavis},
  F.~Guidi\altaffilmark{\UCDavis,\IAP}\orcidlink{0000-0001-7593-3962},
  S.~Guns\altaffilmark{\Berkeley},
  N.~W.~Halverson\altaffilmark{\CASA,\ColoradoPhys},
  E.~Hivon\altaffilmark{\IAP}\orcidlink{0000-0003-1880-2733},
  G.~P.~Holder\altaffilmark{\ILPhys}\orcidlink{0000-0002-0463-6394},
  W.~L.~Holzapfel\altaffilmark{\Berkeley},
  J.~C.~Hood\altaffilmark{\KICPChicago},
  A.~Hryciuk\altaffilmark{\PhysicsUChicago,\KICPChicago},
  N.~Huang\altaffilmark{\Berkeley}\orcidlink{0000-0003-3595-0359},
  T.~Jhaveri\altaffilmark{\AAUChicago,\KICPChicago},
  F.~K\'eruzor\'e\altaffilmark{\ANLHEP},
  A.~R.~Khalife\altaffilmark{\IAP}\orcidlink{0000-0002-8388-4950},
  L.~Knox\altaffilmark{\UCDavis},
  K.~Kornoelje\altaffilmark{\AAUChicago,\KICPChicago,\ANLHEP},
  C.-L.~Kuo\altaffilmark{\KIPAC,\Stanford,\SLAC},
  K.~Levy\altaffilmark{\Melbourne},
  Y.~Li\altaffilmark{\KICPChicago}\orcidlink{0000-0002-4820-1122},
  A.~E.~Lowitz\altaffilmark{\KICPChicago}\orcidlink{0000-0002-4747-4276},
  C.~Lu\altaffilmark{\ILPhys},
  G.~P.~Lynch\altaffilmark{\UCDavis}\orcidlink{0009-0004-3143-1708},
  A.~S.~Maniyar\altaffilmark{\KIPAC,\Stanford,\SLAC}\orcidlink{0000-0002-4617-9320},
  E.~S.~Martsen\altaffilmark{\AAUChicago,\KICPChicago},
  F.~Menanteau\altaffilmark{\ILAst,\ILNCSA},
  M.~Millea\altaffilmark{\Berkeley}\orcidlink{0000-0001-7317-0551},
  J.~Montgomery\altaffilmark{\McGill},
  Y.~Nakato\altaffilmark{\Stanford},
  T.~Natoli\altaffilmark{\KICPChicago},
  A.~Ouellette\altaffilmark{\ILPhys}\orcidlink{0000-0003-0170-5638},
  Z.~Pan\altaffilmark{\ANLHEP,\KICPChicago,\PhysicsUChicago}\orcidlink{0000-0002-6164-9861},
  P.~Paschos\altaffilmark{\EFIChicago},
  K.~A.~Phadke\altaffilmark{\ILAst,\ILNCSA,\SKAI}\orcidlink{0000-0001-7946-557X},
  A.~W.~Pollak\altaffilmark{\AAUChicago},
  K.~Prabhu\altaffilmark{\UCDavis},
  W.~Quan\altaffilmark{\ANLHEP,\PhysicsUChicago,\KICPChicago},
  M.~Rahimi\altaffilmark{\Melbourne},
  A.~Rahlin\altaffilmark{\AAUChicago,\KICPChicago}\orcidlink{0000-0003-3953-1776},
  C.~L.~Reichardt\altaffilmark{\Melbourne}\orcidlink{0000-0003-2226-9169},
  M.~Rouble\altaffilmark{\McGill},
  J.~E.~Ruhl\altaffilmark{\CaseWestern},
  A.~C.~Silva~Oliveira\altaffilmark{\Caltech,\KIPAC,\Stanford}\orcidlink{0000-0001-5755-5865},
  A.~Simpson\altaffilmark{\AAUChicago,\KICPChicago},
  J.~A.~Sobrin\altaffilmark{\FNAL,\KICPChicago}\orcidlink{0000-0001-6155-5315},
  A.~A.~Stark\altaffilmark{\CfA},
  J.~Stephen\altaffilmark{\EFIChicago},
  C.~Trendafilova\altaffilmark{\ILNCSA},
  J.~D.~Vieira\altaffilmark{\ILAst,\ILPhys,\ILNCSA}\orcidlink{0000-0001-7192-3871},
  A.~G.~Vieregg\altaffilmark{\KICPChicago,\AAUChicago,\EFIChicago,\PhysicsUChicago}\orcidlink{0000-0002-4528-9886},
  A.~Vitrier\altaffilmark{\IAP}\orcidlink{0009-0009-3168-092X},
  Y.~Wan\altaffilmark{\ILAst,\ILNCSA},
  N.~Whitehorn\altaffilmark{\MSU}\orcidlink{0000-0002-3157-0407},
  W.~L.~K.~Wu\altaffilmark{\Caltech,\KIPAC,\SLAC}\orcidlink{0000-0001-5411-6920},
  M.~R.~Young\altaffilmark{\FNAL,\KICPChicago},
  and
  J.~A.~Zebrowski\altaffilmark{\KICPChicago,\AAUChicago,\FNAL}
}

\altaffiltext{\ILAst}{Department of Astronomy, University of Illinois Urbana-Champaign, 1002 West Green Street, Urbana, IL, 61801, USA}
\altaffiltext{\PhysicsPrinceton}{Joseph Henry Laboratories of Physics, Jadwin Hall, Princeton University, Princeton, NJ 08544, USA}
\altaffiltext{\TTU}{Department of Physics \& Astronomy, Box 41051, Texas Tech University, Lubbock TX 79409-1051, USA}
\altaffiltext{\FNAL}{Fermi National Accelerator Laboratory, MS209, P.O. Box 500, Batavia, IL, 60510, USA}
\altaffiltext{\KICPChicago}{Kavli Institute for Cosmological Physics, University of Chicago, 5640 South Ellis Avenue, Chicago, IL, 60637, USA}
\altaffiltext{\AAUChicago}{Department of Astronomy and Astrophysics, University of Chicago, 5640 South Ellis Avenue, Chicago, IL, 60637, USA}
\altaffiltext{\Melbourne}{School of Physics, University of Melbourne, Parkville, VIC 3010, Australia}
\altaffiltext{\IAP}{Sorbonne Universit\'e, CNRS, UMR 7095, Institut d'Astrophysique de Paris, 98 bis bd Arago, 75014 Paris, France}
\altaffiltext{\UNM}{Department of Physics and Astronomy, University of New Mexico, Albuquerque, NM, 87131, USA}
\altaffiltext{\ANLHEP}{High-Energy Physics Division, Argonne National Laboratory, 9700 South Cass Avenue, Lemont, IL, 60439, USA}
\altaffiltext{\KIPAC}{Kavli Institute for Particle Astrophysics and Cosmology, Stanford University, 452 Lomita Mall, Stanford, CA, 94305, USA}
\altaffiltext{\Stanford}{Department of Physics, Stanford University, 382 Via Pueblo Mall, Stanford, CA, 94305, USA}
\altaffiltext{\SLAC}{SLAC National Accelerator Laboratory, 2575 Sand Hill Road, Menlo Park, CA, 94025, USA}
\altaffiltext{\LMU}{University Observatory, Faculty of Physics, LMU Munich, Scheinerstr.~1, 81679 Munich, Germany}
\altaffiltext{\EFIChicago}{Enrico Fermi Institute, University of Chicago, 5640 South Ellis Avenue, Chicago, IL, 60637, USA}
\altaffiltext{\PhysicsUChicago}{Department of Physics, University of Chicago, 5640 South Ellis Avenue, Chicago, IL, 60637, USA}
\altaffiltext{\Geneva}{Universit\'e de Gen\`eve, D\'epartement de Physique Th\'eorique, 24 Quai Ansermet, CH-1211 Gen\`eve 4, Switzerland}
\altaffiltext{\NTU}{National Taiwan University, No. 1, Sec. 4, Roosevelt Road, Taipei 106319, Taiwan}
\altaffiltext{\Berkeley}{Department of Physics, University of California, Berkeley, CA, 94720, USA}
\altaffiltext{\Saclay}{Universit\'e Paris-Saclay, Universit\'e Paris Cit\'e, CEA, CNRS, AIM, 91191, Gif-sur-Yvette, France}
\altaffiltext{\KEK}{High Energy Accelerator Research Organization (KEK), Tsukuba, Ibaraki 305-0801, Japan}
\altaffiltext{\McGill}{Department of Physics and McGill Space Institute, McGill University, 3600 Rue University, Montreal, Quebec H3A 2T8, Canada}
\altaffiltext{\CIFAR}{Canadian Institute for Advanced Research, CIFAR Program in Gravity and the Extreme Universe, Toronto, ON, M5G 1Z8, Canada}
\altaffiltext{\USTCAst}{Department of Astronomy, University of Science and Technology of China, Hefei 230026, China}
\altaffiltext{\USTCPhys}{School of Astronomy and Space Science, University of Science and Technology of China, Hefei 230026}
\altaffiltext{\ILPhys}{Department of Physics, University of Illinois Urbana-Champaign, 1110 West Green Street, Urbana, IL, 61801, USA}
\altaffiltext{\UCLA}{Department of Physics and Astronomy, University of California, Los Angeles, CA, 90095, USA}
\altaffiltext{\MSU}{Department of Physics and Astronomy, Michigan State University, East Lansing, MI 48824, USA}
\altaffiltext{\Caltech}{California Institute of Technology, 1200 East California Boulevard., Pasadena, CA, 91125, USA}
\altaffiltext{\UCDavis}{Department of Physics \& Astronomy, University of California, One Shields Avenue, Davis, CA 95616, USA}
\altaffiltext{\CASA}{CASA, Department of Astrophysical and Planetary Sciences, University of Colorado, Boulder, CO, 80309, USA }
\altaffiltext{\ColoradoPhys}{Department of Physics, University of Colorado, Boulder, CO, 80309, USA}
\altaffiltext{\ILNCSA}{Center for AstroPhysical Surveys, National Center for Supercomputing Applications, Urbana, IL, 61801, USA}
\altaffiltext{\SKAI}{NSF-Simons AI Institute for the Sky (SkAI), 172 E. Chestnut St., Chicago, IL 60611, USA}
\altaffiltext{\CaseWestern}{Department of Physics, Case Western Reserve University, Cleveland, OH, 44106, USA}
\altaffiltext{\CfA}{Center for Astrophysics \textbar{} Harvard \& Smithsonian, 60 Garden Street, Cambridge, MA, 02138, USA}